\documentclass[12pt,preprint]{aastex}
\usepackage{natbib}
\shorttitle{Properties of Star Forming Cluster Galaxies}
\shortauthors{Crawford et al.}

\usepackage{lscape}

\begin{document}

\title {Spectroscopy of Luminous Compact Blue Galaxies in Distant Clusters II.
Physical Properties of dE Progenitor Candidates}

\author{S. M. Crawford\footnote{South African Astronomical Observatory,
Observatory, 7935 Cape Town, South Africa; crawford@saao.ac.za}, 
Gregory D. Wirth\footnote{W. M. Keck Observatory, 65-1120 Mamalahoa Hwy, Kamuela HI 96743; gregory.wirth@gmail.com},  
M. A. Bershady,\footnote{Washburn Observatories, U. Wisconsin -
Madison, 475 N. Charter St., Madison, WI 53706; mab@astro.wisc.edu} and
S. M. Randriamampandry\footnote{Astrophysics \& Cosmology Research Unit, School of Mathematics, Statistics \& Computer Science, University of KwaZulu-Natal, Durban 4041, South Africa}
}

\begin{abstract}

 Luminous Compact Blue Galaxies (LCBGs) are an extreme star-bursting  population of galaxies that were far more common at earlier epochs  than today.   Based on spectroscopic and photometric measurements of LCBGs in massive ($M >10^{15} M_{\odot}$), intermediate redshift  ($0.5 < z < 0.9$) galaxy clusters, we present  their rest-frame properties including  star-formation rate, dynamical mass, size, luminosity, and  metallicity. The appearance of these small,  compact galaxies in clusters at intermediate redshift helps explain the observed redshift  evolution in the size-luminosity relationship among cluster  galaxies. In addition, we find  the rest-frame properties of LCBGs appearing in galaxy clusters are indistinguishable from field  LCBGs at the same redshift.   Up to $35\%$ of the LCBGs show significant discrepancies between optical and infrared indicators of star formation, suggesting that star formation occurs in obscured regions.   Nonetheless, the star formation for LCBGs shows a decrease toward the center of the galaxy clusters.   Based on their position and velocity, we estimate that up to 10\% of cluster LCBGs are likely to merge with another cluster galaxy.    Finally, the observed properties and distributions of the LCBGs in these clusters lead us to conclude that we are witnessing the quenching of the progenitors of dwarf elliptical galaxies that dominate the number density of present-epoch galaxy clusters.
  
\end{abstract}


\section{Introduction}

Luminous Compact Blue Galaxies (LCBGs) are an extreme, star-bursting
population of galaxies \citep{1994ApJ...427L...9K}  found both
within clusters of galaxies \citep{1997ApJ...478L..49K, c06} and outside. The number density and
star-formation rate of LCBGs have decreased over time in concert with
the global star formation rate \citep{guzman97}.  However, beyond
having high surface brightness and blue colors, field LCBGs are heterogeneous in their
morphologies: some objects are extremely compact, some
appear to be undergoing nuclear starbursts, and others display very
irregular morphologies \citep{1996ApJ...460L...5G, phillips97,
  1999ApJ...511..118K, 2003ApJ...586L..45G, 2004ApJ...615..689G,
  2004ApJ...617.1004W, 2006ApJ...649..129B,2006ApJ...640L.143N,
  2007A&A...469..483R,2007AJ....134.2455H, 2010ApJ...708.1076T}.  This
heterogenous nature has led researchers to propose a variety of
evolutionary outcomes for LCBGs, including evolving into low-luminosity
spheroidal systems \citep{1994ApJ...427L...9K, 1996ApJ...460L...5G}
and the bulges of spiral galaxies \citep{2001ApJ...550L..35B,
  2001ApJ...550..570H}.

The dramatic increase in the fraction of blue galaxies
\citep{bo84} in clusters at high redshift tracks the increase in
star-forming galaxies in the field \citep{1997ARA&A..35..389E}.  Due
to a variety of cluster processes \citep{bg06}, the blue galaxies in
clusters are likely to have their star formation quenched, and if they
are falling into the cluster for the first time, they will experience
significant stripping of both their gas and stellar material.  For
this reason, star-forming galaxies in clusters have long been proposed
as the possible progenitors of dwarf elliptical galaxies
\citep{1994ApJ...430..121C, 1998ApJ...495..139M, 2005ApJ...619..134T}.

Other than very early work by \cite{1997ApJ...478L..49K} confirming
their presence in clusters, relatively few studies have targeted the
cluster LCBG population; however, recent observations have indicated
that the cluster environment is triggering the LCBG phase \citep{c06,
  c11}.  In fact, their kinematic and spatial distributions suggest
that LCBGs represent galaxies experiencing initial infall into galaxy
clusters \citep{c14}.

In this work, we characterize the properties of LCBGs in galaxy
clusters in order to assess how they might evolve over time.  In
\S  \ref{sec:obs}, we briefly summarize the observational data
underlying this work.  A description of the different techniques used
to derive the rest-frame properties of our sample follows in \S\ref{sec:analysis}.  We present various properties of LCBGs and
compare them to other populations in \S\ref{sec:disc}.  Finally,
in \S\ref{sec:fate}, we present possible fates for LCBGs in galaxy
clusters.  Throughout this work, we adopt $H_0 = 70~\mathrm{km}\ \mathrm{s}^{-1}\ \mathrm{Mpc}^{-1}$, $\Omega_M = 0.3$,
and $\Omega_{\Lambda} = 0.7$.

\subsection{Definition of Galaxy Classifications}

Following \cite{c11, c14}, we use the following definitions for our different samples 
discussed in this paper to divide the galaxy populations into three samples based on their color 
and compactness:

\begin{itemize}

\item Red Sequence (RS) are defined as objects redder than the
  dividing line between blue and red objects, which we define as $U-B
  = -0.032 \times (M_B + 21.52) + 0.204 $ following
  \citet{2006ApJ...647..853W}

\item Blue cloud (BC) galaxies are defined as objects bluer than the
  dividing line between blue and red objects, which we define as $U-B
  = -0.032 \times (M_B + 21.52) + 0.204 $ following
  \citet{2006ApJ...647..853W}

 \item Luminous compact blue galaxies (LCBGs) are a subset of the BC
   class, defined as having $B-V<0.5$, $\mu_B < 21~\mbox{mag
     arcsec}^{-2}$, and $M_B < -18.5$ \citep{2004ApJ...615..689G, c06}.

\end{itemize}

Throughout this work, we compare our results to the field sample produced by \cite{guzman97}.  Their definition
of compact objects is in apparent properties, but once transformed into absolute space, it is very similar to
the definition for LCBGs that we use here.

\section{Observations}\label{sec:obs}

Full details of the observations are presented in our previous works \citep{c06,c11,c14}, but here 
we present a brief description of our sample selection and observations.   The set of galaxy 
clusters used in our work are summarized in Table \ref{tab:all}.  We originally selected 
this set of clusters based on their extreme richness, their intermediate redshifts, and their large
existing set of archive data.  We identified LCBGs through their photometric 
properties from deep imaging on the WIYN telescope for spectroscopic follow-up 
with the  Deep Imaging Multi-Object Spectrograph (DEIMOS) on the Keck Telescope.  Based on their spectroscopic redshifts, LCBGs were identified as
being either part of the cluster or in the field \citep{c14}.   In total, we have 381 LCBGs with
spectroscopic redshifts in our sample.   Of those, 119 are cluster LCBGs and 262 are field LCBGs.  
Of the cluster (field) LCBGs, we have DEIMOS spectra for 53 (80) galaxies and 
all of those galaxies have [\ion{O}{2}] $\lambda3727$  detections.    Of that sample, 33 (41) cluster (field) 
LCBGs have Hubble Space Telescope (HST) imaging and 18 (19) cluster (field) LCBGs have 
far-infrared measurements from the Spitzer Space Telescope.

\begin{landscape}
\begin{deluxetable}{rrrrrrrcr@{\extracolsep{0.1cm}}ccr}
\tabletypesize{\tiny}\tablewidth{0pc}
\tablecaption{Summary of Cluster Observations}
\tablehead{
&
&
&
&
&
\multicolumn{4}{c}{Hubble Space Telescope} & 
\multicolumn{3}{l}{\hspace{1cm}Spitzer Space Telescope} \\

\cline{6-9}  
\cline{10-12}

\colhead{Cluster} & 
\colhead{$\alpha$} &
\colhead{$\delta$} &
\colhead{$z$} &
\colhead{$\sigma_p$} &
\colhead{PropID} & 
\colhead{Instrument} & 
\colhead{ExpTime} & 
\colhead{Filter} & 
\colhead{ExpTime} & 
\colhead{RMS} & 
\colhead{SEIP Data Tag} \\ 

\colhead{} & 
\colhead{(J2000)} &
\colhead{(J2000)} &
\colhead{} & 
\colhead{(km~s$^{-1}$)} &
\colhead{} & 
\colhead{} & 
\colhead{s}  &
\colhead{} & 
\colhead{s} & 
\colhead{$\mu Jy$} & 
\colhead{ADS/IRSA.Atlas/}  
\\
}
\startdata 
\label{tab:all}
MS 0451-03 & 04:54:10.8 & $-$03:00:51 & $     0.5389 $  & $         1328 \pm           47 $ & 9836 & ACS & 2036 & F814W & 1637 & 26 & $\#2013/0325/072424\_12320$ \\  
Cl 0016+16 & 00:18:33.6 & $+$16:26:16 & $     0.5467$  & $         1490 \pm           80 $  & 6825 & WFPC2 & 4600 & F814W  & 764  & 28 & $\#2015/0309/051809\_29746$ \\  
           & & & & & 10493 & ACS & 2183 & F775W & & & \\
Cl~J1324+3011 & 13:24:48.8 & $+$30:11:39 & $     0.7549 $  & $          806 \pm           85 $ &     &      &     &       &     &    &  \\
MS 1054-03 & 10:56:60.0 & $-$03:37:36 & $     0.8307 $  & $         1105 \pm           61 $   & 9772 & ACS & 1980 & F814W & 3218 & 19 & $\#2015/0309/052253\_30941$ \\
          & & & &  & 10196 & ACS & 2160 & F814W & & & \\
Cl J1604+4304 & 16:04:24.0 & $+$43:04:39 & $     0.9005 $  & $         1106 \pm          167 $ & 10496 & ACS &  1500 & F850LP & 1389  & 13 & $\#2015/0309/053231\_935$ \\
\enddata 
\
\tablecomments{(1) Cluster Field of the Observationts (2) and (3)~celestial
  coordinates of the adopted cluster center defined by Brightest
  Cluster Galaxy; (4)~measured cluster redshift; (5)~measured cluster
  projected velocity dispersion (6) HST Proposal ID (7) HST Instrument 
  (8) Typical Exposure Time (9) Filter (10) Mean Exposure Time 
  (11) RMS Noise (12) Spitzer Enhanced Image Product ID}
\end{deluxetable}
\end{landscape}

\subsection{WIYN Imaging}

Deep imaging data in UBRIz and two narrow bands were obtained with the
Mini-Mosaic camera from the WIYN 3.5~m telescope for all five clusters
between 1999 October and 2004 June.  The Mini-Mosaic camera has a
$9\farcm6\times9\farcm6$ FOV with $0\farcs14~\mathrm{px}^{-1}$
plate scale.  In the $R$ band, the median value for the seeing was
FWHM=$0\farcs85$.  The narrow-band filters were specifically designed
to detect [\ion{O}{2}] $\lambda$3727 at the redshift of each cluster.
Details of the data reduction and analysis appear in \cite{c09}.

\subsection{Spectroscopic Observations}

We obtained spectroscopic observations for a sample of cluster
star-forming galaxies with the Deep Imaging Multi-Object Spectrograph
\citep[DEIMOS,][]{2003SPIE.4841.1657F} on the Keck~II Telescope during
October 2005 and April 2007.  Different instrument configurations were
used in order to sample key diagnostic features at the redshift of
each cluster.  Full details of the observations and data reductions
are given in \cite{c11}.

\subsection{HST Imaging}

All of the clusters have been the target of extensive observations with the Hubble Space Telescope (HST), primarily using either WFPC2 or the Adva nced Camera for Surveys (ACS).  WFPC2 has a
chevron-shaped field of view over $2\farcm5\times2\farcm5$ region with
a pixel scale of $0\farcs1~\mathrm{px}^{-1}$ \citep{wfpc2}.  The ACS
camera has a field of view of $3\farcm7\times3\farcm7$ with a pixel
scale of $0\farcs05~\mathrm{px}^{-1}$ \citep{acs}. For all
measurements, we have attempted to select data taken in a filter
closest to the rest-frame $B$ band.  We have employed ACS imaging data
whenever possible and substituted WFPC2 images only when required.  In Table~\ref{tab:all}, we summarize all of the HST observations that are used
in this work.

\subsection{Spitzer Observations}

For clusters observed in the far-infrared regime by the Spitzer Space Telescope, we extracted MIPS 24 \micron \  flux densities, $S_{24}$, from images obtained through the Enhanced Imaging Products\footnote{\texttt{http://irsa.ipac.caltech.edu/data/SPITZER/Enhanced/SEIP/}} archive following the procedures outlined in 
 \cite{2015MNRAS.447..168R}.  In Table~\ref{tab:all}, we list the
 details of each of these observations.

\section{Data Analysis}\label{sec:analysis}

In this section, we describe the different methods we used to
determine the properties of the galaxies, including measuring their
photometry, size, rest-frame properties, and spectroscopic properties.
In Table~\ref{tab:source}, we report on the properties of these
sources.  

\begin{landscape}
\begin{deluxetable}{rccclllllll}
\tabletypesize{\tiny}\tablewidth{0pc}
\tablecaption{Measured Properties of LCBGs}
\label{tab:source}
\tablehead{
\colhead{Object} & 
\colhead{$\alpha$} & 
\colhead{$\beta$} & 
\colhead{z} & 
\colhead{M$_B$} &
\colhead{r$_e$} &
\colhead{$\sigma$} &
\colhead{Mass} &
\colhead{SFR(OII)} &
\colhead{SFR(IR)} &
\colhead{12+log(O/H)}
\\ 
\colhead{} & 
\colhead{(J2000)} & 
\colhead{(J2000)} & 
\colhead{} & 
\colhead{(mag)}  &
\colhead{(kpc)}  &
\colhead{($km \ s^{-1}$)} & 
\colhead{($10^9 \ M_{\odot}$)} &
\colhead{($M_{\odot} \ yr^{-1}$)} &
\colhead{($M_{\odot} \ yr^{-1}$)} &
\colhead{(dex)} 
\\
\colhead{(1)} & 
\colhead{(2)} & 
\colhead{(3)} & 
\colhead{(4)} & 
\colhead{(5)} & 
\colhead{(6)} & 
\colhead{(7)} & 
\colhead{(8)} & 
\colhead{(9)} & 
\colhead{(10)} & 
\colhead{(11)}  \\

}
\startdata 
\multicolumn{10}{c}{Cluster LCBG} \\
\hline

WLTV  J001828.43+162615.7 &  4.6184760 & 16.4377106 & 0.5442 & $-18.57\pm0.05$ & $3.33 \pm 1.65^H$ & $   26 \pm  3$ & $ 2.5 \pm 1.3$ & $0.54 \pm 0.2$ & ... & $8.7 \pm 0.8$ \\ 
WLTV  J001829.15+162653.4 &  4.6214648 & 16.4481822 & 0.5378 & $-19.65\pm0.04$ & $1.89 \pm 0.29^H$ & $   90 \pm  5$ & $17.4 \pm 2.8$ & $0.97 \pm 0.1$ & $14.09 \pm 0.4$ &$7.8 \pm 0.4$ \\ 
WLTV  J001831.90+162441.1 &  4.6329173 & 16.4114185 & 0.5464 & $-19.99\pm0.04$ & $2.27 \pm 0.38^H$ & $   59 \pm  9$ & $ 9.0 \pm 2.1$ & $0.35 \pm 0.0$ & ... & ... \\ 
WLTV  J001834.81+162719.9 &  4.6450259 & 16.4555305 & 0.5392 & $-20.12\pm0.04$ & $2.65 \pm 0.64^W$ & $   68 \pm 13$ & $13.9 \pm 4.4$ & $3.26 \pm 0.3$ & $4.68 \pm 0.9$ &$8.4 \pm 0.4$ \\ 
WLTV  J001836.15+162635.5 &  4.6506133 & 16.4431949 & 0.5469 & $-20.26\pm0.04$ & $1.29 \pm 0.65^W$ & $   44 \pm  9$ & $ 2.8 \pm 1.6$ & $0.83 \pm 0.1$ & ... & $8.9 \pm 0.3$ \\ 
WLTV  J001836.69+162645.6 &  4.6528944 & 16.4460093 & 0.5571 & $-19.35\pm0.05$ & $2.63 \pm 0.88^H$ & $  110 \pm 12$ & $36.2 \pm 12.7$ & $0.59 \pm 0.1$ & ... & $9.0 \pm 0.4$ \\ 
WLTV  J001839.47+162414.2 &  4.6644431 & 16.4039527 & 0.5559 & $-20.19\pm0.05$ & $1.30 \pm 0.65^W$ & $   45 \pm  2$ & $ 3.0 \pm 1.5$ & $2.91 \pm 0.8$ & ... & $9.0 \pm 0.5$ \\ 
WLTV  J001838.51+162513.7 &  4.6604379 & 16.4204798 & 0.5371 & $-19.77\pm0.04$ & $2.58 \pm 0.51^H$ & $   49 \pm 13$ & $ 7.0 \pm 2.3$ & $0.72 \pm 0.1$ & ... & $8.4 \pm 0.3$ \\ 
WLTV  J001841.98+162524.8 &  4.6749359 & 16.4235667 & 0.5573 & $-20.04\pm0.06$ & $1.90 \pm 0.21^H$ & $   23 \pm  2$ & $ 1.1 \pm 0.2$ & $1.66 \pm 0.7$ & ... & $8.9 \pm 1.0$ \\ 
WLTV  J001843.56+162606.0 &  4.6815191 & 16.4350089 & 0.5541 & $-20.57\pm0.04$ & $3.34 \pm 0.01^H$ & $   48 \pm  2$ & $ 8.9 \pm 0.5$ & $1.17 \pm 0.0$ & ... & $9.0 \pm 0.1$ \\ 
WLTV  J001845.83+162643.3 &  4.6909432 & 16.4453646 & 0.5393 & $-21.21\pm0.03$ & $1.75 \pm 0.03^H$ & $   50 \pm  4$ & $ 5.0 \pm 0.5$ & $4.70 \pm 0.1$ & $15.53 \pm 0.4$ &$8.8 \pm 0.0$ \\ 
WLTV  J001845.90+162748.4 &  4.6912518 & 16.4634667 & 0.5394 & $-19.46\pm0.04$ & $1.60 \pm 0.65^H$ & $   36 \pm 14$ & $ 2.4 \pm 1.3$ & $1.78 \pm 0.3$ & ... & $8.4 \pm 0.6$ \\ 
WLTV  J001847.60+162614.8 &  4.6983474 & 16.4374512 & 0.5528 & $-20.14\pm0.04$ & $2.14 \pm 0.65^W$ & $   32 \pm  2$ & $ 2.6 \pm 0.8$ & $1.05 \pm 0.2$ & $11.34 \pm 0.5$ &$9.0 \pm 0.3$ \\ 
WLTV  J045358.78-025857.2 & 73.4949225 & -2.9825687 & 0.5381 & $-20.83\pm0.04$ & $3.82 \pm 1.08^H$ & $   26 \pm  2$ & $ 3.0 \pm 0.9$ & $1.38 \pm 0.2$ & ... & ... \\ 
WLTV  J045401.42-030125.6 & 73.5059232 & -3.0238011 & 0.5401 & $-20.10\pm0.04$ & $2.10 \pm 0.18^H$ & $   37 \pm  3$ & $ 3.4 \pm 0.5$ & $2.88 \pm 0.2$ & $6.88 \pm 0.6$ &$8.5 \pm 0.2$ \\ 
WLTV  J045401.78-030054.4 & 73.5074178 & -3.0151296 & 0.5321 & $-20.19\pm0.04$ & $2.78 \pm 0.02^H$ & $   66 \pm  7$ & $13.8 \pm 1.6$ & $2.71 \pm 0.2$ & $10.11 \pm 0.4$ &$8.6 \pm 0.1$ \\ 
WLTV  J045401.86-030029.2 & 73.5077568 & -3.0081267 & 0.5114 & $-20.86\pm0.03$ & $1.31 \pm 0.03^H$ & $   53 \pm  3$ & $ 4.2 \pm 0.3$ & $5.26 \pm 0.1$ & $6.71 \pm 0.6$ &$8.7 \pm 0.0$ \\ 
WLTV  J045403.40-025920.3 & 73.5141490 & -2.9889974 & 0.5315 & $-19.74\pm0.04$ & $2.33 \pm 0.47^H$ & $   72 \pm  8$ & $13.8 \pm 3.2$ & $1.42 \pm 0.3$ & $20.51 \pm 0.2$ &$8.7 \pm 0.3$ \\ 
WLTV  J045405.10-025939.1 & 73.5212652 & -2.9942039 & 0.5271 & $-22.13\pm0.03$ & $6.84 \pm 0.18^H$ & $   49 \pm  3$ & $18.4 \pm 1.3$ & $1.53 \pm 0.5$ & $23.35 \pm 0.2$ &... \\ 
WLTV  J045405.56-025918.3 & 73.5231756 & -2.9884231 & 0.5300 & $-22.03\pm0.04$ & $6.73 \pm 0.23^H$ & $  184 \pm  5$ & $255.1 \pm 12.1$ & $3.40 \pm 0.1$ & $17.61 \pm 0.3$ &$8.7 \pm 0.0$ \\ 
WLTV  J045405.27-025950.0 & 73.5219589 & -2.9972233 & 0.5388 & $-20.35\pm0.04$ & $2.40 \pm 0.12^H$ & $   74 \pm 26$ & $14.8 \pm 5.3$ & $1.26 \pm 0.1$ & ... & $8.7 \pm 0.1$ \\ 
WLTV  J045406.22-030020.3 & 73.5259274 & -3.0056456 & 0.5311 & $-19.03\pm0.04$ & $1.63 \pm 0.04^H$ & $   50 \pm 24$ & $ 4.6 \pm 2.2$ & $0.97 \pm 0.2$ & ... & $8.6 \pm 0.4$ \\ 
WLTV  J045411.85-030407.7 & 73.5493753 & -3.0688174 & 0.5292 & $-20.24\pm0.04$ & $3.08 \pm 0.23^H$ & $  108 \pm  8$ & $40.8 \pm 4.4$ & $0.30 \pm 0.0$ & $14.78 \pm 0.4$ &... \\ 
WLTV  J045412.07-030201.2 & 73.5502925 & -3.0336742 & 0.5285 & $-19.77\pm0.04$ & $2.36 \pm 0.36^H$ & $   42 \pm  3$ & $ 4.7 \pm 0.8$ & $1.27 \pm 0.1$ & ... & $7.9 \pm 0.3$ \\ 
WLTV  J045411.58-025630.4 & 73.5482314 & -2.9417774 & 0.5470 & $-18.98\pm0.06$ & $2.52 \pm 0.95^H$ & $   26 \pm  2$ & $ 2.0 \pm 0.8$ & $1.97 \pm 1.5$ & ... & ... \\ 
WLTV  J045412.68-030054.2 & 73.5528434 & -3.0150628 & 0.5443 & $-20.75\pm0.04$ & $1.87 \pm 0.10^H$ & $   47 \pm 28$ & $ 4.8 \pm 2.8$ & $0.84 \pm 0.0$ & $15.09 \pm 0.3$ &$8.9 \pm 0.1$ \\ 
WLTV  J045413.51-030148.4 & 73.5562996 & -3.0301159 & 0.5478 & $-20.19\pm0.04$ & $3.12 \pm 0.41^H$ & $   77 \pm 28$ & $20.7 \pm 8.0$ & $1.35 \pm 0.1$ & $4.24 \pm 0.8$ &$8.9 \pm 0.1$ \\ 
WLTV  J045414.98-030424.4 & 73.5624305 & -3.0734561 & 0.5299 & $-20.30\pm0.04$ & $3.40 \pm 0.33^H$ & $   80 \pm  4$ & $24.7 \pm 2.8$ & $1.53 \pm 0.1$ & ... & $8.5 \pm 0.2$ \\ 
WLTV  J045415.18-030331.9 & 73.5632546 & -3.0588815 & 0.5351 & $-20.04\pm0.04$ & $1.85 \pm 0.16^H$ & $   44 \pm  2$ & $ 4.2 \pm 0.4$ & $2.66 \pm 0.2$ & ... & $8.6 \pm 0.2$ \\ 
WLTV  J045421.25-030325.9 & 73.5885575 & -3.0571959 & 0.5420 & $-19.38\pm0.04$ & $2.26 \pm 0.36^H$ & $<  10$ & $ 0.2 \pm 0.1$ & $0.75 \pm 0.1$ & ... & $8.1 \pm 0.6$ \\ 
WLTV  J045421.31-030230.0 & 73.5888003 & -3.0416920 & 0.5109 & $-18.64\pm0.05$ & $2.29 \pm 0.55^H$ & $   45 \pm  4$ & $ 5.2 \pm 1.4$ & $0.51 \pm 0.1$ & ... & $8.7 \pm 0.5$ \\ 
WLTV  J045424.11-030348.0 & 73.6004433 & -3.0633382 & 0.5062 & $-19.92\pm0.04$ & $2.08 \pm 0.12^H$ & $   62 \pm  7$ & $ 9.2 \pm 1.2$ & $1.05 \pm 1.0$ & ... & $8.0 \pm 1.3$ \\ 
WLTV  J105655.78-033544.2 & 164.2324078 & -3.5956237 & 0.8244 & $-18.88\pm0.04$ & $0.91 \pm 0.08^H$ & $   69 \pm  4$ & $ 4.9 \pm 0.5$ & $0.90 \pm 0.4$ & ... & ... \\ 
WLTV  J105706.06-033603.6 & 164.2752425 & -3.6010122 & 0.8293 & $-19.54\pm0.06$ & $2.17 \pm 0.42^H$ & $<  10$ & $ 0.2 \pm 0.0$ & $1.06 \pm 0.5$ & $3.32 \pm 0.7$ &... \\ 
WLTV  J105712.03-034135.3 & 164.3001342 & -3.6931475 & 0.8319 & $-18.56\pm0.07$ & $1.67 \pm 0.77^W$ & $   44 \pm 15$ & $ 3.7 \pm 2.2$ & $0.40 \pm 0.4$ & ... & ... \\ 
WLTV  J105653.85-033402.1 & 164.2243902 & -3.5672541 & 0.8305 & $-20.72\pm0.05$ & $2.17 \pm 0.42^H$ & $   29 \pm 12$ & $ 2.2 \pm 1.0$ & ... & ... & ... \\ 
WLTV  J132445.28+301031.0 & 201.1886774 & 30.1752830 & 0.7529 & $-19.81\pm0.06$ & $3.54 \pm 0.74^W$ & $  148 \pm  7$ & $87.5 \pm 18.9$ & $0.92 \pm 0.1$ & ... & $8.8 \pm 0.3$ \\ 
WLTV  J132448.33+300748.5 & 201.2013649 & 30.1301409 & 0.7613 & $-20.90\pm0.04$ & $4.98 \pm 0.75^W$ & $   91 \pm  8$ & $46.1 \pm 8.2$ & $0.80 \pm 0.2$ & ... & $9.1 \pm 0.4$ \\ 
WLTV  J132526.34+300926.7 & 201.3597635 & 30.1574307 & 0.7530 & $-20.17\pm0.05$ & $2.74 \pm 0.74^W$ & $  136 \pm  5$ & $56.6 \pm 15.5$ & $1.41 \pm 0.1$ & ... & $8.9 \pm 0.4$ \\ 
WLTV  J132527.34+301002.5 & 201.3639153 & 30.1673657 & 0.7414 & $-20.17\pm0.05$ & $3.88 \pm 0.74^W$ & $  140 \pm 29$ & $86.0 \pm 24.3$ & $0.69 \pm 0.1$ & ... & ... \\ 
WLTV  J132531.77+301119.8 & 201.3823760 & 30.1888353 & 0.7417 & $-20.15\pm0.04$ & $3.51 \pm 0.74^W$ & $   50 \pm  3$ & $10.1 \pm 2.3$ & $1.85 \pm 0.2$ & ... & $8.8 \pm 1.3$ \\ 
WLTV  J160407.17+430444.2 & 241.0298930 & 43.0789545 & 0.9010 & $-22.54\pm0.09$ & $4.91 \pm 0.79^W$ & $  120 \pm 15$ & $80.1 \pm 16.6$ & $5.05 \pm 0.6$ & ... & $9.2 \pm 0.7$ \\ 
WLTV  J160410.04+430400.6 & 241.0418319 & 43.0668580 & 0.9085 & $-20.58\pm0.05$ & $5.10 \pm 0.79^W$ & $   55 \pm  7$ & $17.2 \pm 3.5$ & $4.56 \pm 1.0$ & ... & ... \\ 
WLTV  J160410.74+430420.5 & 241.0447607 & 43.0723872 & 0.9083 & $-21.61\pm0.04$ & $7.27 \pm 0.79^W$ & $   37 \pm  3$ & $11.3 \pm 1.7$ & $4.17 \pm 0.3$ & $3.90 \pm 0.4$ &$8.8 \pm 1.2$ \\ 
WLTV  J160418.72+430720.8 & 241.0779907 & 43.1224440 & 0.9047 & $-18.92\pm0.08$ & $2.11 \pm 0.79^W$ & $   64 \pm 92$ & $ 9.7 \pm 14.4$ & $0.84 \pm 0.2$ & ... & ... \\ 
WLTV  J160418.99+430331.9 & 241.0791221 & 43.0588682 & 0.8820 & $-21.12\pm0.08$ & $1.65 \pm 0.85^H$ & $  190 \pm 49$ & $67.0 \pm 38.7$ & $2.92 \pm 0.4$ & $24.25 \pm 0.1$ &$8.9 \pm 0.4$ \\ 
WLTV  J160423.09+430208.8 & 241.0962242 & 43.0357862 & 0.8944 & $-21.30\pm0.05$ & $6.17 \pm 0.79^W$ & $   83 \pm  7$ & $48.2 \pm 7.6$ & $8.80 \pm 0.8$ & $0.77 \pm 0.9$ &$8.3 \pm 1.5$ \\ 
WLTV  J160425.18+430416.8 & 241.1049144 & 43.0713390 & 0.9010 & $-21.29\pm0.04$ & $5.34 \pm 2.50^H$ & $   42 \pm  5$ & $11.0 \pm 5.3$ & $2.97 \pm 0.3$ & $9.27 \pm 0.2$ &$9.2 \pm 0.2$ \\ 
WLTV  J160447.47+430757.4 & 241.1977973 & 43.1326179 & 0.9037 & $-20.38\pm0.07$ & $2.93 \pm 0.79^W$ & $   47 \pm  5$ & $ 7.2 \pm 2.2$ & $3.89 \pm 0.5$ & ... & $6.7 \pm 0.4$ \\ 
WLTV  J160447.88+430805.8 & 241.1994986 & 43.1349592 & 0.9023 & $-22.07\pm0.04$ & $3.50 \pm 0.79^W$ & $   64 \pm  6$ & $16.1 \pm 3.9$ & $16.47 \pm 0.6$ & ... & $7.0 \pm 0.3$ \\ 
\hline
\multicolumn{10}{c}{Field LCBG} \\
\hline
WLTV  J001814.36+162511.0 &  4.5598337 & 16.4197472 & 0.6556 & $-21.33\pm0.04$ & $5.66 \pm 0.70^W$ & $   91 \pm  6$ & $52.5 \pm 7.5$ & $2.31 \pm 0.1$ & $27.16 \pm 0.3$ &$9.0 \pm 0.1$ \\ 
WLTV  J001824.52+162713.9 &  4.6021678 & 16.4538719 & 0.5825 & $-20.48\pm0.04$ & $1.79 \pm 0.13^H$ & $   73 \pm  4$ & $10.7 \pm 1.0$ & $1.44 \pm 0.2$ & $8.91 \pm 0.5$ &$9.0 \pm 0.2$ \\ 
WLTV  J001827.44+162606.7 &  4.6143268 & 16.4352003 & 0.4865 & $-18.65\pm0.04$ & $1.77 \pm 0.40^H$ & $   76 \pm  3$ & $11.7 \pm 2.7$ & $0.73 \pm 1.6$ & ... & ... \\ 
WLTV  J001830.64+162642.9 &  4.6276554 & 16.4452699 & 0.6533 & $-19.83\pm0.05$ & $2.35 \pm 0.80^H$ & $   23 \pm  3$ & $ 1.5 \pm 0.6$ & $2.00 \pm 1.2$ & ... & ... \\ 
WLTV  J001834.30+162648.9 &  4.6429167 & 16.4469257 & 0.6926 & $-19.76\pm0.04$ & $2.56 \pm 0.62^H$ & $   78 \pm  7$ & $17.5 \pm 4.6$ & $1.48 \pm 0.3$ & $18.32 \pm 0.3$ &$8.2 \pm 0.2$ \\ 
WLTV  J001835.19+162658.8 &  4.6466418 & 16.4496750 & 0.4858 & $-18.95\pm0.05$ & $1.85 \pm 0.40^H$ & $<  10$ & $ 0.2 \pm 0.1$ & $0.35 \pm 0.5$ & ... & $8.6 \pm 1.8$ \\ 
WLTV  J001835.82+162611.4 &  4.6492322 & 16.4365057 & 0.6227 & $-19.95\pm0.06$ & $1.37 \pm 0.69^W$ & $  109 \pm 13$ & $18.2 \pm 9.4$ & $4.92 \pm 2.1$ & ... & ... \\ 
WLTV  J001844.66+162459.9 &  4.6860672 & 16.4166487 & 0.8330 & $-20.58\pm0.04$ & $4.33 \pm 1.23^H$ & $  102 \pm 55$ & $50.5 \pm 31.2$ & $3.24 \pm 0.5$ & $10.48 \pm 0.5$ &$7.4 \pm 0.4$ \\ 
WLTV  J001846.64+162547.3 &  4.6943428 & 16.4298197 & 0.4384 & $-20.36\pm0.04$ & $1.54 \pm 0.01^H$ & $   42 \pm  1$ & $ 3.1 \pm 0.1$ & $2.32 \pm 0.1$ & $4.49 \pm 0.8$ &$8.8 \pm 0.0$ \\ 
WLTV  J001819.48+162533.1 &  4.5811615 & 16.4258787 & 0.8269 & $-20.84\pm0.05$ & $9.11 \pm 3.49^H$ & $   54 \pm 12$ & $30.0 \pm 13.3$ & $8.58 \pm 5.3$ & ... & $5.8 \pm 0.7$ \\ 
WLTV  J045354.78-030203.9 & 73.4782672 & -3.0344341 & 0.7190 & $-20.12\pm0.04$ & $1.94 \pm 0.05^H$ & $  126 \pm  7$ & $34.9 \pm 2.2$ & $3.37 \pm 0.6$ & ... & ... \\ 
WLTV  J045358.38-025618.5 & 73.4932513 & -2.9384845 & 0.8255 & $-20.06\pm0.04$ & $2.96 \pm 0.64^H$ & $   66 \pm 68$ & $14.4 \pm 15.3$ & $2.69 \pm 0.4$ & ... & $6.9 \pm 0.4$ \\ 
WLTV  J045359.94-030306.1 & 73.4997697 & -3.0516944 & 0.5679 & $-19.65\pm0.05$ & $2.62 \pm 0.33^H$ & $   96 \pm 12$ & $27.1 \pm 4.8$ & $0.67 \pm 0.3$ & $12.55 \pm 0.5$ &... \\ 
WLTV  J045400.93-030312.6 & 73.5038636 & -3.0535020 & 0.5697 & $-21.43\pm0.04$ & $2.96 \pm 0.14^H$ & $   54 \pm  3$ & $ 9.7 \pm 0.8$ & $3.12 \pm 2.2$ & $27.45 \pm 0.2$ &... \\ 
WLTV  J045402.65-025918.4 & 73.5110531 & -2.9884637 & 0.7150 & $-19.61\pm0.04$ & $1.21 \pm 0.23^H$ & $   43 \pm  4$ & $ 2.5 \pm 0.5$ & $1.04 \pm 0.2$ & ... & $7.8 \pm 0.4$ \\ 
WLTV  J045405.49-030202.4 & 73.5228918 & -3.0340243 & 0.5689 & $-18.90\pm0.05$ & $1.04 \pm 0.20^H$ & $   48 \pm  4$ & $ 2.8 \pm 0.6$ & $0.38 \pm 0.1$ & ... & ... \\ 
WLTV  J045407.09-025837.4 & 73.5295585 & -2.9770594 & 0.5674 & $-20.03\pm0.04$ & $1.23 \pm 0.12^H$ & $   29 \pm  5$ & $ 1.2 \pm 0.2$ & $1.03 \pm 0.7$ & ... & ... \\ 
WLTV  J045407.76-030055.5 & 73.5323133 & -3.0154195 & 0.5677 & $-20.93\pm0.04$ & $4.34 \pm 0.47^H$ & $  118 \pm  4$ & $67.6 \pm 7.7$ & $2.79 \pm 0.1$ & ... & ... \\ 
WLTV  J045409.67-030232.2 & 73.5403107 & -3.0423026 & 0.3545 & $-19.98\pm0.04$ & $2.12 \pm 0.07^H$ & $   15 \pm  2$ & $ 0.5 \pm 0.1$ & $1.62 \pm 0.1$ & $6.36 \pm 0.6$ &$8.3 \pm 0.1$ \\ 
WLTV  J045409.67-030043.0 & 73.5402878 & -3.0119636 & 0.4439 & $-18.65\pm0.04$ & $2.48 \pm 1.33^H$ & $   29 \pm  3$ & $ 2.5 \pm 1.3$ & $0.31 \pm 0.1$ & $3.71 \pm 0.8$ &$7.4 \pm 1.3$ \\ 
WLTV  J045411.02-030125.6 & 73.5459121 & -3.0238001 & 0.5896 & $-20.11\pm0.04$ & $1.20 \pm 0.05^H$ & $   35 \pm  1$ & $ 1.7 \pm 0.1$ & $3.15 \pm 0.4$ & $10.90 \pm 0.4$ &$8.8 \pm 0.2$ \\ 
WLTV  J045410.97-025913.8 & 73.5457235 & -2.9871760 & 0.9865 & $-21.36\pm0.04$ & $0.31 \pm 0.02^H$ & $   62 \pm 10$ & $ 1.4 \pm 0.3$ & $15.77 \pm 1.7$ & ... & $5.4 \pm 0.4$ \\ 
WLTV  J045412.21-030328.2 & 73.5508668 & -3.0578372 & 0.6548 & $-20.63\pm0.04$ & $1.40 \pm 0.70^W$ & $   87 \pm 10$ & $11.9 \pm 6.2$ & $1.83 \pm 0.2$ & $16.54 \pm 0.3$ &$9.0 \pm 0.4$ \\ 
WLTV  J045412.10-030325.0 & 73.5504014 & -3.0572215 & 0.6554 & $-20.43\pm0.04$ & $3.52 \pm 0.52^H$ & $   89 \pm 30$ & $31.2 \pm 11.7$ & $1.58 \pm 0.2$ & $19.75 \pm 0.3$ &$8.9 \pm 0.3$ \\ 
WLTV  J045415.68-030411.9 & 73.5653136 & -3.0699730 & 0.7760 & $-20.45\pm0.04$ & $2.66 \pm 0.30^H$ & $   36 \pm  5$ & $ 4.0 \pm 0.7$ & $1.82 \pm 0.2$ & $6.66 \pm 0.6$ &$8.1 \pm 0.4$ \\ 
WLTV  J045420.19-030346.8 & 73.5841290 & -3.0630249 & 0.6186 & $-20.52\pm0.04$ & $1.93 \pm 0.13^H$ & $   69 \pm  3$ & $10.4 \pm 0.8$ & $3.15 \pm 0.3$ & ... & $8.6 \pm 0.5$ \\ 
WLTV  J045420.57-030517.2 & 73.5856996 & -3.0881327 & 0.8183 & $-20.77\pm0.04$ & $2.77 \pm 0.24^H$ & $   56 \pm  5$ & $ 9.7 \pm 1.2$ & $1.69 \pm 0.3$ & $13.63 \pm 0.4$ &$8.4 \pm 0.4$ \\ 
WLTV  J045422.66-030503.1 & 73.5944373 & -3.0842073 & 0.5799 & $-20.23\pm0.04$ & $2.62 \pm 0.22^H$ & $   80 \pm  5$ & $19.0 \pm 2.0$ & $1.83 \pm 0.3$ & ... & $8.9 \pm 0.2$ \\ 
WLTV  J045422.73-030416.3 & 73.5947091 & -3.0712078 & 0.8848 & $-20.37\pm0.05$ & $2.17 \pm 0.32^H$ & $  146 \pm  7$ & $52.0 \pm 8.1$ & $2.57 \pm 0.4$ & $14.93 \pm 0.6$ &$7.5 \pm 0.4$ \\ 
WLTV  J045424.70-030303.6 & 73.6029358 & -3.0510237 & 0.9860 & $-20.49\pm0.04$ & $4.34 \pm 0.81^W$ & $   62 \pm 42$ & $19.1 \pm 13.4$ & $5.57 \pm 1.2$ & ... & $6.1 \pm 0.4$ \\ 
WLTV  J045425.76-030513.9 & 73.6073246 & -3.0871954 & 0.7330 & $-20.76\pm0.04$ & $3.44 \pm 0.05^H$ & $   69 \pm  7$ & $18.8 \pm 2.1$ & $4.76 \pm 0.4$ & ... & $7.0 \pm 0.3$ \\ 
WLTV  J045427.31-025831.5 & 73.6137908 & -2.9754424 & 0.5778 & $-18.71\pm0.05$ & $2.00 \pm 0.48^H$ & $   38 \pm  5$ & $ 3.3 \pm 0.9$ & $0.76 \pm 0.4$ & ... & $8.7 \pm 2.0$ \\ 
WLTV  J132435.56+301124.8 & 201.1481469 & 30.1902250 & 0.7815 & $-20.19\pm0.04$ & $1.50 \pm 0.75^W$ & $   73 \pm  6$ & $ 9.1 \pm 4.6$ & $2.18 \pm 0.1$ & ... & $7.5 \pm 0.3$ \\ 
WLTV  J132446.14+301020.9 & 201.1922365 & 30.1724843 & 0.7015 & $-20.15\pm0.04$ & $3.16 \pm 0.72^W$ & $   79 \pm 13$ & $22.2 \pm 6.3$ & $1.07 \pm 0.2$ & ... & ... \\ 
WLTV  J132456.78+301159.2 & 201.2365773 & 30.1997992 & 0.5940 & $-18.77\pm0.05$ & $2.38 \pm 0.67^W$ & $   74 \pm  9$ & $14.9 \pm 4.6$ & $0.01 \pm 0.0$ & ... & $9.0 \pm 1.8$ \\ 
WLTV  J132457.29+301157.8 & 201.2387127 & 30.1994066 & 0.7998 & $-19.71\pm0.04$ & $3.12 \pm 0.76^W$ & $   46 \pm  5$ & $ 7.4 \pm 2.0$ & $1.56 \pm 0.5$ & ... & $7.3 \pm 0.5$ \\ 
WLTV  J132458.04+301115.5 & 201.2418274 & 30.1876456 & 0.7849 & $-19.54\pm0.04$ & $3.49 \pm 0.76^W$ & $   41 \pm 10$ & $ 6.8 \pm 2.3$ & $1.30 \pm 0.2$ & ... & $8.9 \pm 0.7$ \\ 
WLTV  J132458.78+301136.3 & 201.2449283 & 30.1934218 & 0.7998 & $-19.50\pm0.04$ & $3.22 \pm 0.76^W$ & $   45 \pm  5$ & $ 7.4 \pm 1.9$ & $1.06 \pm 0.2$ & ... & $6.6 \pm 1.4$ \\ 
WLTV  J132500.25+300853.2 & 201.2510585 & 30.1481219 & 0.6180 & $-20.85\pm0.04$ & $3.68 \pm 0.69^W$ & $   75 \pm  2$ & $23.7 \pm 4.5$ & $0.09 \pm 0.1$ & ... & $9.1 \pm 1.0$ \\ 
WLTV  J132505.84+301200.6 & 201.2743179 & 30.2001930 & 0.8383 & $-20.12\pm0.05$ & $3.17 \pm 0.77^W$ & $   66 \pm 11$ & $15.7 \pm 4.6$ & $1.52 \pm 0.2$ & ... & $7.9 \pm 0.4$ \\ 
WLTV  J132509.33+301010.6 & 201.2888767 & 30.1696218 & 0.6070 & $-21.06\pm0.04$ & $3.73 \pm 0.68^W$ & $   50 \pm  3$ & $10.5 \pm 2.0$ & $0.10 \pm 0.1$ & ... & $9.2 \pm 1.0$ \\ 
WLTV  J132515.80+301111.8 & 201.3158279 & 30.1866204 & 0.8402 & $-21.26\pm30.00$ & $1.54 \pm 0.94^W$ & $   59 \pm  6$ & $ 6.1 \pm 3.8$ & $5.29 \pm 0.3$ & ... & $7.7 \pm 0.3$ \\ 
WLTV  J132520.16+301107.7 & 201.3340026 & 30.1854901 & 0.6591 & $-20.92\pm0.04$ & $3.35 \pm 0.71^W$ & $   75 \pm  1$ & $21.1 \pm 4.5$ & $7.66 \pm 0.1$ & ... & $8.7 \pm 0.0$ \\ 
WLTV  J132524.05+301021.6 & 201.3502129 & 30.1726779 & 0.6597 & $-19.75\pm0.04$ & $2.89 \pm 0.71^W$ & $   93 \pm  6$ & $27.9 \pm 7.1$ & $0.03 \pm 0.0$ & ... & $9.0 \pm 1.0$ \\ 
WLTV  J132524.70+301213.2 & 201.3529109 & 30.2036743 & 0.6075 & $-20.78\pm0.04$ & $3.14 \pm 0.68^W$ & $   79 \pm  3$ & $22.3 \pm 4.9$ & $0.08 \pm 0.1$ & ... & $9.0 \pm 1.0$ \\ 
WLTV  J132529.04+300734.6 & 201.3710158 & 30.1262793 & 0.9203 & $-18.79\pm0.04$ & $1.58 \pm 0.80^W$ & $   63 \pm  5$ & $ 7.0 \pm 3.6$ & $0.60 \pm 0.0$ & ... & $7.5 \pm 0.3$ \\ 
WLTV  J132538.84+301048.2 & 201.4118496 & 30.1800576 & 0.6979 & $-20.86\pm0.04$ & $3.52 \pm 0.72^W$ & $   62 \pm  3$ & $15.6 \pm 3.3$ & $0.09 \pm 0.1$ & ... & $9.3 \pm 1.1$ \\ 
WLTV  J160350.90+430654.6 & 240.9620764 & 43.1151916 & 0.8314 & $-19.46\pm0.11$ & $1.53 \pm 0.78^W$ & $   43 \pm  3$ & $ 3.2 \pm 1.6$ & $0.02 \pm 0.0$ & ... & ... \\ 
WLTV  J160351.75+430748.7 & 240.9656222 & 43.1302128 & 0.8610 & $-20.14\pm0.05$ & $4.50 \pm 0.78^W$ & $   54 \pm  5$ & $14.9 \pm 3.0$ & $2.50 \pm 0.5$ & ... & $8.6 \pm 1.9$ \\ 
WLTV  J160408.43+430009.8 & 241.0351107 & 43.0027345 & 0.5798 & $-19.87\pm0.04$ & $3.54 \pm 0.66^W$ & $   43 \pm  3$ & $ 7.6 \pm 1.6$ & $0.03 \pm 0.0$ & $1.59 \pm 0.5$ &$9.3 \pm 1.1$ \\ 
WLTV  J160415.93+430618.1 & 241.0663830 & 43.1050541 & 0.8290 & $-19.16\pm0.06$ & $2.66 \pm 0.77^W$ & $   34 \pm  3$ & $ 3.6 \pm 1.1$ & $0.02 \pm 0.0$ & ... & ... \\ 
WLTV  J160422.15+430542.5 & 241.0922999 & 43.0951607 & 0.8300 & $-21.05\pm0.04$ & $4.48 \pm 1.57^H$ & $   72 \pm  5$ & $26.2 \pm 9.4$ & $4.63 \pm 0.4$ & ... & ... \\ 
WLTV  J160427.90+430401.9 & 241.1162686 & 43.0672186 & 0.8275 & $-19.28\pm0.09$ & $1.99 \pm 1.10^H$ & $  124 \pm 20$ & $34.1 \pm 19.8$ & $0.86 \pm 0.9$ & $15.89 \pm 0.1$ &... \\ 
WLTV  J160428.02+430007.4 & 241.1167467 & 43.0020768 & 0.7441 & $-22.08\pm0.06$ & $1.87 \pm 0.74^W$ & $   57 \pm  4$ & $ 6.8 \pm 2.8$ & $0.26 \pm 0.3$ & ... & ... \\ 
WLTV  J160429.08+425955.0 & 241.1211631 & 42.9986169 & 0.7507 & $-18.97\pm0.06$ & $3.03 \pm 0.74^W$ & $   52 \pm  6$ & $ 9.3 \pm 2.6$ & $0.02 \pm 0.0$ & $6.23 \pm 0.2$ &$9.1 \pm 1.9$ \\ 
WLTV  J160433.16+430452.7 & 241.1381739 & 43.0813248 & 0.8665 & $-18.80\pm0.08$ & $2.81 \pm 0.29^H$ & $   34 \pm  6$ & $ 3.7 \pm 0.8$ & $0.58 \pm 0.6$ & ... & ... \\ 
WLTV  J160442.65+430827.8 & 241.1777033 & 43.1410584 & 0.6203 & $-19.68\pm0.05$ & $2.71 \pm 0.68^W$ & $   40 \pm  4$ & $ 4.9 \pm 1.3$ & $0.03 \pm 0.0$ & $12.80 \pm 0.2$ &... \\ 
WLTV  J160429.42+425945.7 & 241.1225975 & 42.9960265 & 0.5882 & $-20.47\pm0.04$ & $2.88 \pm 0.67^W$ & $   45 \pm  2$ & $ 6.6 \pm 1.6$ & $0.06 \pm 0.1$ & ... & $9.0 \pm 1.0$ \\ 
WLTV  J160446.27+431023.5 & 241.1927985 & 43.1732114 & 0.6948 & $-19.34\pm0.74$ & $1.92 \pm 0.72^W$ & $   32 \pm  1$ & $ 2.2 \pm 0.8$ & $0.02 \pm 0.0$ & ... & $8.6 \pm 1.5$ \\ 
\hline
\multicolumn{10}{c}{Cluster Blue Galaxy} \\
\hline
WLTV  J001813.32+162440.8 &  4.5555066 & 16.4113337 & 0.5450 & $-21.29\pm0.04$ & $9.34 \pm 0.65^W$ & ... & ... & $1.61 \pm 0.1$ & ... & $8.7 \pm 0.1$ \\ 
WLTV  J001813.61+162455.0 &  4.5567216 & 16.4153018 & 0.5454 & $-20.89\pm0.04$ & $4.72 \pm 0.64^W$ & ... & ... & $1.42 \pm 0.1$ & ... & $9.0 \pm 0.1$ \\ 
WLTV  J001814.69+162426.2 &  4.5612138 & 16.4073028 & 0.5491 & $-21.43\pm0.04$ & $8.07 \pm 0.65^W$ & ... & ... & $1.39 \pm 0.1$ & ... & $9.0 \pm 0.1$ \\ 
WLTV  J001816.48+162547.2 &  4.5686686 & 16.4297923 & 0.5467 & $-19.17\pm0.06$ & $7.15 \pm 0.65^W$ & ... & ... & $1.76 \pm 1.5$ & ... & $8.6 \pm 1.0$ \\ 
WLTV  J001819.21+162415.6 &  4.5800526 & 16.4043495 & 0.5463 & $-19.47\pm0.04$ & $4.03 \pm 0.89^H$ & ... & ... & $0.85 \pm 0.3$ & ... & $8.6 \pm 0.9$ \\ 
WLTV  J001820.09+162318.9 &  4.5836895 & 16.3885837 & 0.5728 & $-20.64\pm0.04$ & $2.58 \pm 0.14^H$ & ... & ... & $1.15 \pm 0.2$ & ... & ... \\ 
WLTV  J001822.44+162531.6 &  4.5934932 & 16.4254592 & 0.5442 & $-20.51\pm0.04$ & $17.20 \pm 0.67^W$ & ... & ... & $1.15 \pm 0.5$ & ... & ... \\ 
WLTV  J001823.68+162502.6 &  4.5986727 & 16.4174012 & 0.5516 & $-18.83\pm0.06$ & $1.47 \pm 0.59^H$ & ... & ... & $0.67 \pm 0.5$ & ... & $9.0 \pm 1.4$ \\ 
WLTV  J001824.22+162513.2 &  4.6008981 & 16.4203332 & 0.5526 & $-19.49\pm0.04$ & $1.68 \pm 0.26^H$ & ... & ... & $1.10 \pm 0.1$ & ... & $8.9 \pm 0.1$ \\ 
WLTV  J001836.54+162515.1 &  4.6522550 & 16.4208647 & 0.5532 & $-20.02\pm0.04$ & $3.46 \pm 0.67^H$ & ... & ... & $0.10 \pm 0.0$ & ... & ... \\ 
WLTV  J001840.20+162506.6 &  4.6674911 & 16.4185034 & 0.5381 & $-20.19\pm0.04$ & $3.09 \pm 0.03^H$ & ... & ... & $1.40 \pm 0.2$ & ... & $7.9 \pm 0.2$ \\ 
WLTV  J001845.29+162706.5 &  4.6886946 & 16.4518150 & 0.5385 & $-21.08\pm0.04$ & $7.16 \pm 0.64^W$ & ... & ... & $7.48 \pm 0.7$ & ... & $8.6 \pm 0.2$ \\ 
WLTV  J001848.15+162556.1 &  4.7006305 & 16.4322493 & 0.5325 & $-19.57\pm0.05$ & $5.34 \pm 0.64^W$ & ... & ... & $0.59 \pm 0.2$ & ... & ... \\ 
WLTV  J045358.36-030126.6 & 73.4931775 & -3.0240815 & 0.5369 & $-19.82\pm0.05$ & $4.43 \pm 0.10^H$ & ... & ... & $0.88 \pm 0.1$ & ... & $8.4 \pm 0.2$ \\ 
WLTV  J045358.88-025828.0 & 73.4953229 & -2.9744572 & 0.5081 & $-18.38\pm0.05$ & $2.15 \pm 0.62^W$ & ... & ... & $0.18 \pm 0.2$ & ... & $8.6 \pm 1.9$ \\ 
WLTV  J045359.56-025803.5 & 73.4981627 & -2.9676489 & 0.5511 & $-20.84\pm0.04$ & $5.65 \pm 0.55^H$ & ... & ... & $1.16 \pm 0.1$ & ... & ... \\ 
WLTV  J045400.13-030207.8 & 73.5005414 & -3.0355113 & 0.5482 & $-18.44\pm0.05$ & $1.63 \pm 0.39^H$ & ... & ... & $0.65 \pm 0.2$ & ... & $8.4 \pm 0.7$ \\ 
WLTV  J045401.41-025859.8 & 73.5058703 & -2.9833001 & 0.5314 & $-19.58\pm0.04$ & $2.96 \pm 1.42^H$ & ... & ... & $0.68 \pm 0.1$ & ... & $8.7 \pm 0.3$ \\ 
WLTV  J045404.50-030013.6 & 73.5187366 & -3.0037808 & 0.5322 & $-21.41\pm0.04$ & $7.39 \pm 0.38^H$ & ... & ... & $2.31 \pm 0.1$ & ... & $8.8 \pm 0.0$ \\ 
WLTV  J045406.77-030031.4 & 73.5282151 & -3.0087413 & 0.5453 & $-20.35\pm0.04$ & $3.84 \pm 0.91^H$ & ... & ... & $1.32 \pm 0.2$ & ... & $8.9 \pm 0.2$ \\ 
WLTV  J045415.86-030356.1 & 73.5660694 & -3.0656041 & 0.5260 & $-20.78\pm0.04$ & $5.05 \pm 0.48^H$ & ... & ... & $3.12 \pm 0.1$ & ... & $8.9 \pm 0.1$ \\ 
WLTV  J045428.96-030506.8 & 73.6206506 & -3.0852374 & 0.5086 & $-19.60\pm0.04$ & $4.24 \pm 1.06^H$ & ... & ... & $0.73 \pm 0.1$ & ... & $7.7 \pm 0.7$ \\ 
WLTV  J105638.95-034041.1 & 164.1623084 & -3.6781093 & 0.8298 & $-19.51\pm0.08$ & $6.07 \pm 0.77^W$ & ... & ... & $1.16 \pm 0.8$ & ... & ... \\ 
WLTV  J105644.75-033348.8 & 164.1864420 & -3.5635784 & 0.8340 & $-19.19\pm0.09$ & $5.85 \pm 2.54^H$ & ... & ... & $0.68 \pm 0.2$ & ... & ... \\ 
WLTV  J105702.54-033948.1 & 164.2605863 & -3.6633828 & 0.8320 & $-20.28\pm0.07$ & $5.26 \pm 0.77^W$ & ... & ... & $1.88 \pm 1.9$ & ... & ... \\ 
WLTV  J105704.45-033515.9 & 164.2685356 & -3.5877566 & 0.8268 & $-19.35\pm0.06$ & $2.76 \pm 0.89^H$ & ... & ... & $1.35 \pm 0.5$ & ... & ... \\ 
WLTV  J105707.27-033538.6 & 164.2802861 & -3.5940554 & 0.8410 & $-20.16\pm0.04$ & $1.79 \pm 0.59^H$ & ... & ... & $1.51 \pm 0.4$ & ... & ... \\ 
WLTV  J105708.46-033512.3 & 164.2852546 & -3.5867494 & 0.8374 & $-20.12\pm0.06$ & $4.46 \pm 1.05^H$ & ... & ... & $0.94 \pm 0.4$ & ... & ... \\ 
WLTV  J105711.46-033540.5 & 164.2977662 & -3.5945981 & 0.8392 & $-16.52\pm0.07$ & $2.54 \pm 0.06^H$ & ... & ... & $0.04 \pm 0.0$ & ... & ... \\ 
WLTV  J105713.91-033548.1 & 164.3079645 & -3.5967120 & 0.8307 & $-20.20\pm0.09$ & $2.00 \pm 0.77^W$ & ... & ... & $1.69 \pm 0.5$ & ... & ... \\ 
WLTV  J132436.19+300843.1 & 201.1507784 & 30.1453114 & 0.7528 & $-20.48\pm0.04$ & $5.72 \pm 0.74^W$ & ... & ... & $4.58 \pm 0.3$ & ... & $8.7 \pm 0.3$ \\ 
WLTV  J132440.55+301043.1 & 201.1689704 & 30.1786405 & 0.7552 & $-20.25\pm0.05$ & $6.62 \pm 0.75^W$ & ... & ... & $1.27 \pm 0.1$ & ... & ... \\ 
WLTV  J132440.31+301114.7 & 201.1679695 & 30.1874284 & 0.7580 & $-18.99\pm0.06$ & $3.16 \pm 0.75^W$ & ... & ... & $0.64 \pm 0.2$ & ... & ... \\ 
WLTV  J132442.46+300922.8 & 201.1769086 & 30.1563440 & 0.7562 & $-18.84\pm0.08$ & $3.73 \pm 0.75^W$ & ... & ... & $0.64 \pm 0.1$ & ... & ... \\ 
WLTV  J132448.20+301103.2 & 201.2008414 & 30.1842277 & 0.7621 & $-20.94\pm0.04$ & $7.60 \pm 0.75^W$ & ... & ... & $1.10 \pm 0.1$ & ... & $7.5 \pm 0.6$ \\ 
WLTV  J132449.26+301115.6 & 201.2052417 & 30.1876712 & 0.7493 & $-21.14\pm0.04$ & $1.48 \pm 0.74^W$ & ... & ... & $2.63 \pm 0.2$ & ... & $8.8 \pm 0.5$ \\ 
WLTV  J132450.90+300822.4 & 201.2120723 & 30.1395759 & 0.7575 & $-18.30\pm0.08$ & $2.53 \pm 0.75^W$ & ... & ... & $0.37 \pm 0.6$ & ... & ... \\ 
WLTV  J132501.60+301147.2 & 201.2566720 & 30.1964475 & 0.7538 & $-20.60\pm0.04$ & $5.05 \pm 0.74^W$ & ... & ... & $3.35 \pm 0.2$ & ... & $8.7 \pm 0.3$ \\ 
WLTV  J132504.26+301140.8 & 201.2677703 & 30.1946767 & 0.7557 & $-20.09\pm0.05$ & $7.89 \pm 0.75^W$ & ... & ... & $1.01 \pm 0.2$ & ... & $8.7 \pm 1.3$ \\ 
WLTV  J132513.12+300906.3 & 201.3046708 & 30.1517526 & 0.7369 & $-20.91\pm0.06$ & $7.03 \pm 0.74^W$ & ... & ... & $4.16 \pm 0.1$ & ... & $8.4 \pm 0.2$ \\ 
WLTV  J132537.93+300844.1 & 201.4080236 & 30.1455895 & 0.7210 & $-19.22\pm0.05$ & $1.46 \pm 0.73^W$ & ... & ... & $0.02 \pm 0.0$ & ... & ... \\ 
WLTV  J160354.27+430744.8 & 240.9761376 & 43.1291212 & 0.8742 & $-18.68\pm0.06$ & $4.18 \pm 0.78^W$ & ... & ... & $0.68 \pm 0.3$ & ... & ... \\ 
WLTV  J160409.68+430248.6 & 241.0403379 & 43.0468537 & 0.9102 & $-21.82\pm0.06$ & $9.56 \pm 0.79^W$ & ... & ... & $3.49 \pm 1.1$ & ... & ... \\ 
WLTV  J160409.73+430711.9 & 241.0405555 & 43.1199890 & 0.8985 & $-21.39\pm0.08$ & $16.30 \pm 0.80^W$ & ... & ... & $13.56 \pm 11.0$ & ... & ... \\ 
WLTV  J160410.03+430358.4 & 241.0417980 & 43.0662410 & 0.9089 & $-20.50\pm0.05$ & $16.91 \pm 0.80^W$ & ... & ... & $3.94 \pm 4.6$ & ... & ... \\ 
WLTV  J160410.69+430430.3 & 241.0445465 & 43.0750950 & 0.9069 & $-19.47\pm0.05$ & $5.26 \pm 0.79^W$ & ... & ... & $1.96 \pm 0.4$ & ... & ... \\ 
WLTV  J160417.27+430654.0 & 241.0719602 & 43.1150006 & 0.9085 & $-16.33\pm0.06$ & $7.27 \pm 0.79^W$ & ... & ... & $0.10 \pm 0.1$ & ... & ... \\ 
WLTV  J160431.37+430616.8 & 241.1307060 & 43.1046845 & 0.8960 & $-18.93\pm0.09$ & $3.86 \pm 0.79^W$ & ... & ... & $0.49 \pm 0.2$ & ... & $9.1 \pm 1.2$ \\ 
WLTV  J160431.98+430554.4 & 241.1332349 & 43.0984543 & 0.9173 & $-19.77\pm0.07$ & $3.24 \pm 0.80^W$ & ... & ... & $2.18 \pm 0.4$ & ... & $6.7 \pm 0.4$ \\ 
WLTV  J160432.07+430713.3 & 241.1336345 & 43.1203664 & 0.9002 & $-21.35\pm0.06$ & $6.92 \pm 0.79^W$ & ... & ... & $6.36 \pm 8.9$ & ... & ... \\ 
WLTV  J160433.57+430813.8 & 241.1398813 & 43.1371722 & 0.8982 & $-18.59\pm0.09$ & $4.30 \pm 0.79^W$ & ... & ... & $0.59 \pm 0.1$ & ... & ... \\ 
\enddata 
\
\tablecomments{(1) Identification in WLTV survey (2) Right Ascension (3) Declination (4) redshift (5) Absolute B-band magnitude (6)  Half-light radius (7) Velocity dispersion (8) Dynamical Mass (9) SFR from  OII$\lambda3727$ (10) SFR from 24 \micron m flux (11) Oxygen Abundance}
\end{deluxetable}
\end{landscape}

\subsection{Photometry}

We measured aperture and total magnitudes for each source from the ground-based and HST imaging data following the procedure described in \cite{c09}.  For the aperture photometry measurements using the ground-based data only,  we convolved the ground-based images with a 2-D Gaussian function to match the seeing of the worst band for that cluster so that the apertures would be sampling the same light profiles in the different bands.   To define aperture magnitudes, we set the photometric aperture to a diameter of 7.5 kpc at the redshift of the cluster, the size corresponding to the average disk scale length of an $L^*$ galaxy.

For the total magnitudes, we use the same methodology as in our previous work \citep{c06,c09}.  We measured the total magnitude within an aperture that would enclose 99\% of the light based on the profile shape of the source.  The profile shape of the source is determined by measuring the concentration index,
$C_{2080}$.  Following \cite{1984ApJS...56..105K} and \cite{2000AJ....119.2645B}, we define $C_{2080}$ as
\begin{equation}
C_{2080} = 5 \log(r_{20}/r_{80})
\end{equation}
where $r_n$ is the radius that encloses $n\%$ of the light.

\subsection{Size}

We derived the half-light radius ($r_{50}$) for each source from the curve of growth 
we constructed in measuring the total magnitudes. We  measured the radius in 
both the ground-based and HST images. 

\begin{figure}
  \epsscale{0.9}
  \plotone{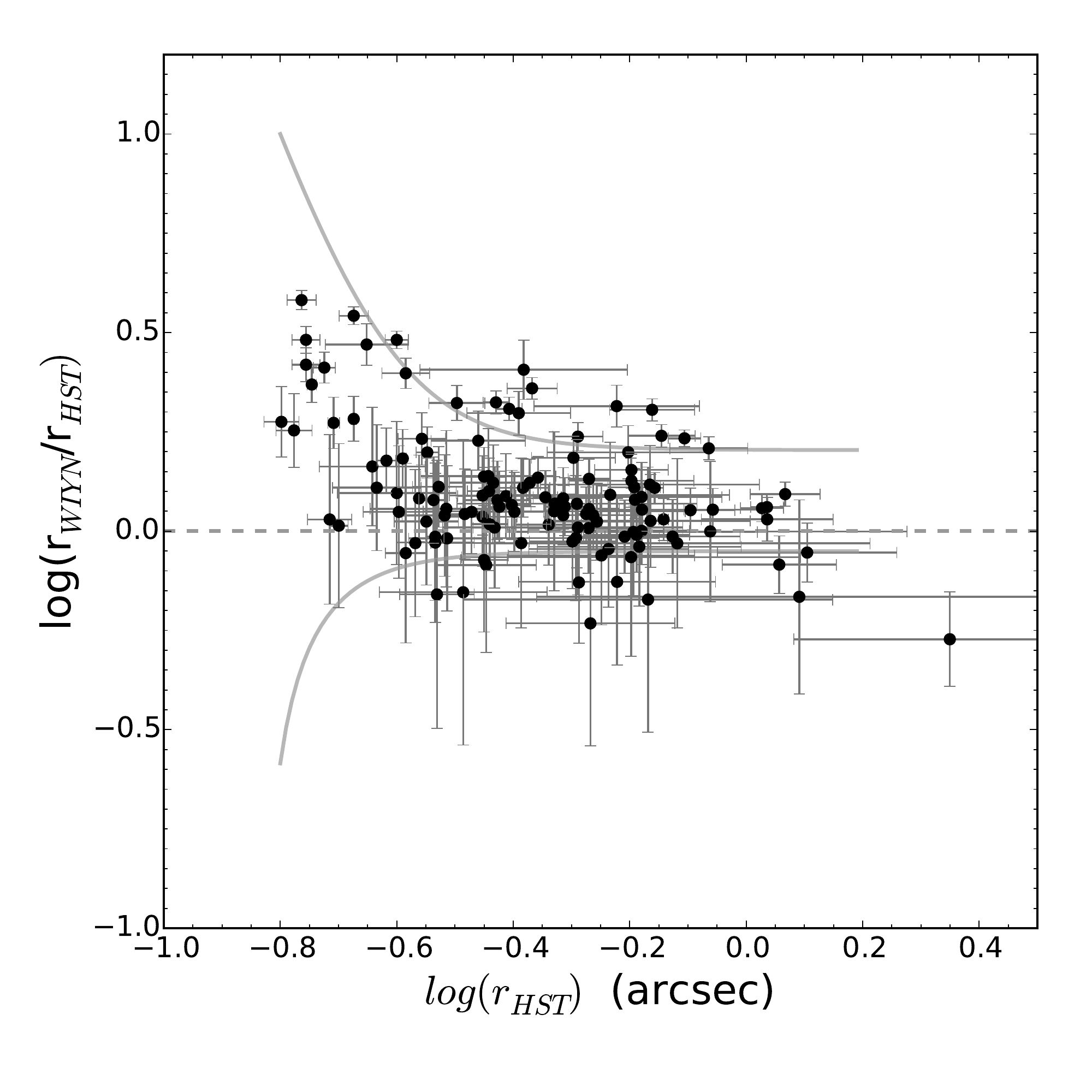}
  \caption{ \label{fig:size}  Comparison between corrected half-light measurements from WIYN observations and HST observations for LCBGs as a function of HST half-light radius.   The corrected WIYN $r_e$ values are consistent with the HST measurements with a median difference of  $0\farcs07$.  This is consistent with a proportional offset of 20\% and a rms scatter of 25\% within the critical half-light size range of 0.3 to 3 arcsec. Solid lines approximate this offset and scatter. The precision and accuracy is more than adequate for our characterization of LCBG size and surface-brightness. } 
\end{figure}

Although the HST images offer superior spatial resolution, they do not
cover a wide projected area and hence miss many objects in our
survey. In contrast, the lower-resolution WIYN images permit us to
measure sizes for more objects.  We used the sizes measured
independently from WIYN and HST to establish the accuracy of the sizes
derived from the ground-based images.  A simple quadrature subtraction
of the seeing value as measured from stellar objects in the WIYN
frames returns a reasonable measurement of the size, as depicted in
Figure \ref{fig:size}.  The median difference between the WIYN
measurements and the HST measurements is $0\farcs065$, although
significant scatter exists at values less than $0\farcs3$.

Whenever possible, we used the half-light radius as measured in the
HST data.  We have not rejected any objects based on the source of
their size measurements.  The provenance of the size measurement is 
indicated in the table.

\subsection{Rest-Frame Properties}

For each of our sources, we determined the absolute $B$-band
magnitude, the rest-frame colors of the sources, and the size.  To
determine the rest-frame magnitude and colors, we fit the spectral
energy distribution (SED) of each galaxy to a range of artificial
templates generated from \textsc{galev} \citep{bc03} and empirical galaxy
templates.  To calculate $M_B$, we derive the K-correction for that
model between the rest-frame $B$ band and the passband closest to the
redshift $B$ band.  For example, for a galaxy at $z=0.54$, the
rest-frame $B$ band is closest to the observed $R$ band, and so the
K-correction is calculated and applied to the apparent total $R$-band
magnitude.  A similar process is used for the rest-frame colors, but
we use the aperture magnitudes instead of the total magnitudes. We applied the 
appropriate cosmological corrections in converting angular measures 
of size to absolute radii.

\subsection{Velocity Width}

The superb wavelength resolution ($10~\mathrm{km~s}^{-1}$ ) of the Keck DEIMOS spectra permit us to measure the velocity widths of even the narrowest of the galaxies in our sample based on their emission lines, greatly enhancing the utility of our study
For each spectroscopic source, we fit either single or double Gaussian
functions to the emission lines associated with [\ion{O}{2}] $\lambda$3727, H$\beta$, and [\ion{O}{3}] $\lambda$5007.  We performed the fit after after subtracting off a low-order polynomial fit to the
continuum of the spectrum near the feature of interest. The DEIMOS spectra resolved the [\ion{O}{2}] $\lambda$3727 doublet into two components, allowing us to fit a double Gaussian with the line 
separation fixed in the rest frame but the remaining parameters allowed to vary.
We visually inspected all fits to confirm  quality.  

To correct for instrumental effects, we follow the procedure from
\cite{guzman97} and subtract, in quadrature, the instrumental
dispersion from the measured value to recover the intrinsic velocity
dispersion of the galaxy.  For most sources, we estimated the
instrumental dispersion based on measurements of nearby sky lines.
For one mask (\texttt{w05.m2}), the seeing during the observations was
exceptional and compact sources did not fill the slit uniformly.  For
compact targets on this mask, the velocity dispersion for all lines
was well below the value measured for the sky lines in the spectra.
For these sources, we estimated their image size based on their spatial
extent in the slit and corrected the dispersion of the sky lines
assuming the effective slit size was equivalent to the full-width half
maximum of the image size.  Similar undersampling problems could
affect additional masks in our survey, so for our most compact sources
we may be underestimating the velocity dispersion.  In addition, we
did not correct the fits for absorption or for any observed rotation
or irregularities in the line profiles.   In the worst case, this would introduce 
an uncertainty of approximately a factor of $\sim2$.   Out of all our spectroscopic 
sources, only $\sim25\%$ showed visual evidence of rotation.

After correcting for the instrumental effects, we determined the final
velocity width of each source by calculating the weighted average of
the measurements for the three lines.  The weights are based on the
inverse variance for each of the lines.  The average velocity
dispersion from all of the lines is reported in
Table~\ref{tab:source}.  Errors for each source were based on
combining errors from each line in quadrature.  We estimate that the smallest
velocity dispersion that we can safely recover is $10~\mathrm{km~s}^{-1}$ 
based on the instrumental velocity dispersion. Any value below this is reported 
as an upper limit.

\subsection{Galaxy Mass}

Using the average velocity width and best size measurement, we
calculated the dynamical mass of each system following
\cite{phillips97} as:
\begin{equation}
  M_{dyn}= \frac{3 c_2}{G} \sigma_v^2 r_e,
\end{equation}
where we take c2, a geometric factor, to be 1.6.   This equation is based 
on an assumption of a virialized system, but, to within a factor of 1.4, it should give a similar 
result as for a rotational supported system \citep{phillips97} .

In addition, we correct the velocity dispersion for a factor of 1.3 following \cite{2003ApJ...586L..45G} to correct for the difference between measuring emission lines instead of stellar absorption lines.   When our data lack sufficient resolution to determine the velocity width and size of certain sources, we report only upper limits on dynamical mass in Table \ref{tab:source}.

\subsection{Equivalent Width}

We followed two methods to measure the equivalent width (EW).  The
first method calculates the continuum level within two regions lying
outside of the line of interest, and then computes the sum over a
region that includes the line of interest to estimate the line flux.
The summed region was designed to include all of the line flux while
avoiding any of the line flux in the continuum measurements.
The second method is based on integrating the Gaussian function fit to
all of our emission lines while measuring the velocity width (see the
previous sections for details of the velocity width measurement).  The lines that
we measured include [\ion{O}{2}] $\lambda3727$, H$\beta$, and [\ion{O}{3}] $\lambda5007$.
Both techniques yield comparable results with a standard deviation of  15\%. We
adopt the first method for our EW values presented here.

\subsection{Star-Formation Rate}

For comparison with \cite{guzman97}, we follow the process outlined in
their appendix for converting [\ion{O}{2}] $\lambda3727$ EW to
star-formation rate (SFR).  We apply the following formula to our data
to estimate the SFR:
\begin{equation}
  \textrm{SFR} (M_{\odot}~\mathrm{yr}^{-1}) = 2.5 \times 10^{-0.4(M_B - M_{B\odot})} \mathrm{EW}_{3727}
\end{equation}
where $M_B$ is the absolute $B$-band magnitude and $\mathrm{EW}_{3727}$
is the [\ion{O}{2}] $\lambda3727$ EW.  This equation also includes a
factor for the average extinction in blue galaxies.   The SFR is a factor of
$\sim 3$ lower compared to the  H$\alpha$ calibration from \cite{1992ApJ...388..310K}
due to the more "top-heavy" IMF in  \cite{1996MNRAS.278..417A} with an upper mass limit 
of $M<125 \ M_{\odot}$  compared to $M<100 \ M_{\odot}$ in the previous work.  
Values for the SFR are listed in Table~\ref{tab:source}.

In addition to the SFR calculated from [\ion{O}{2}] $\lambda3727$, we
also calculate the $\mathrm{SFR}_{\mathrm{FIR}}$ based on the flux from the Spitzer
24~$\micron$ flux.  We follow the method outlined in
\cite{2009ApJ...692..556R} and as applied in
\cite{2015MNRAS.447..168R} for converting the 24~$\micron$ flux to SFR
based on their Equation~14 and Table~1.

\subsection{Metallicity Measurements}

We adopt the method of \cite{2004ApJ...617..240K} for measuring the
metallicity of our objects.  This variation on the $R_{23}$ method
\citep{1979MNRAS.189...95P} uses the ratio of equivalent widths as
opposed to flux ratios.  Similar to other works, we corrected our
H$\beta$ EW by adding a value of 2~\AA\ to it to account for stellar
absorption and our total [\ion{O}{3}] contribution was given by $1.3
\times \mathrm{EW}_{5007}$.  From the values for
$R_{23}$ and $O_{23}$, we then used their Eq.~18 to calculate the
$12+\log[\mathrm{O/H}]$ metallicities for our galaxies.  Unfortunately, degeneracies between the lines available for measuring the metallicity prevent us from distinguishing between the higher and
lower metallicity tracks, but as in previous studies 
or star-forming sources \citep{2004ApJ...617..240K} , we assume that
LCBGs lie on the  upper track with metallicities above $12+\log[\mathrm{O/H}]=8.4$.
Regardless, we make the same
assumptions for cluster and field sources so that comparison
between the two are consistent.

\section{Properties of LCBGs}\label{sec:disc}

\subsection{Magnitude-Size Relationship}

\begin{figure}[p]
  \epsscale{0.9}
  \plotone{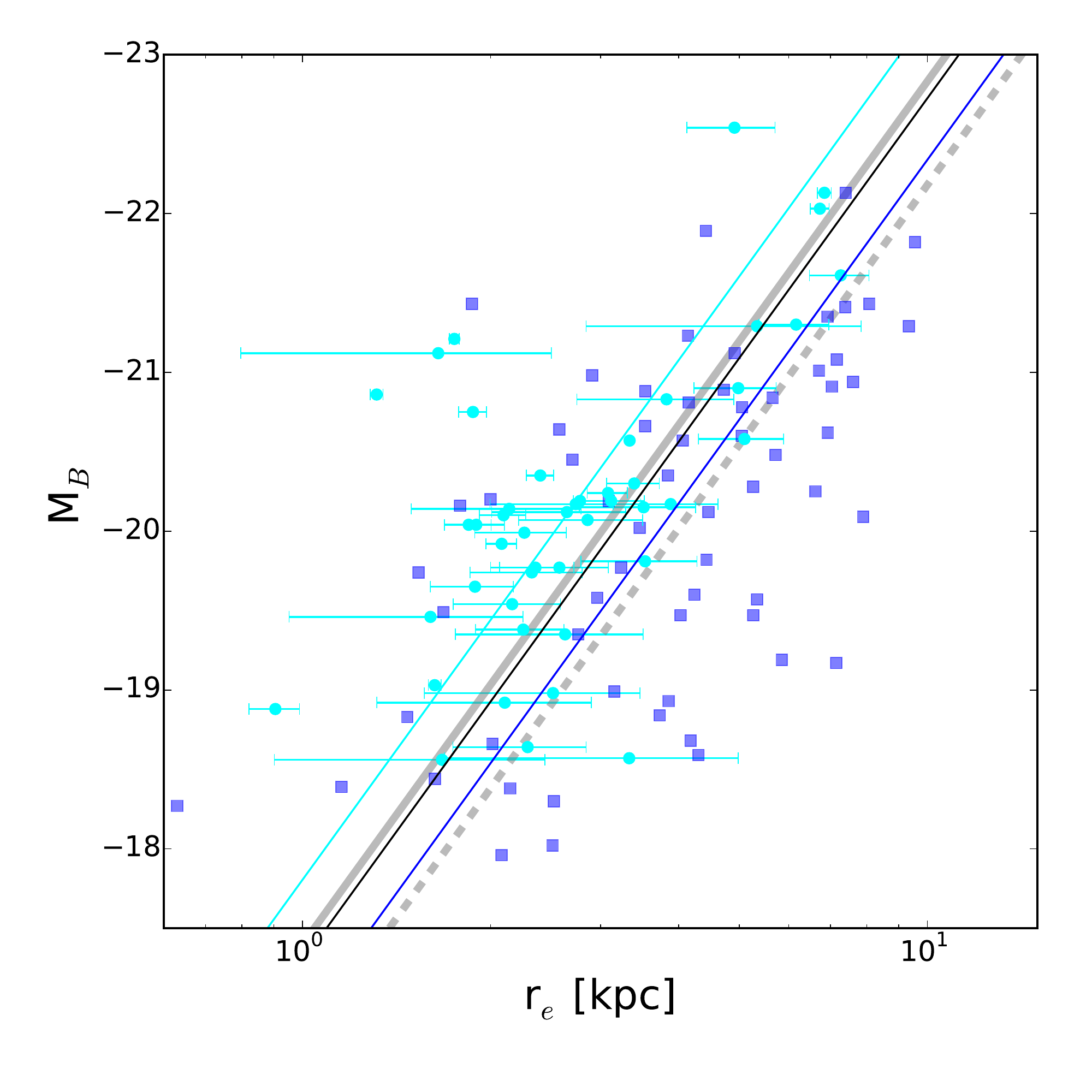}
  \caption{
    \label{fig:mbsize}
    Magnitude-size relationship for cluster emission-line galaxies in
    our sample.  Cluster LCBGs are represented as filled turquoise
    circles and cluster BC galaxies as filled blue squares.  In addition,
    we plot the relationship for disk galaxies found by
    \cite{bamford07} in the low-redshift field (dashed, grey line) and
    in intermediate clusters (solid, grey line).  Solid lines also
    represent the best-fit values for all of our sources (black line),
    LCBGs (turquoise line), and non-LCBG galaxies (blue line).   }
\end{figure}

In Figure~\ref{fig:mbsize}, we plot the relationship between 
magnitude and size for LCBGs and Blue Cloud (BC) galaxies.  For
comparison, we also plot the relationships found by \cite{bamford07} for 
low-redshift disk galaxies (dashed grey line) and intermediate redshift
disk galaxies (solid grey line) in clusters.  The first and most noticeable
feature in the plot is that the LCBG galaxies exhibit smaller
sizes at a given luminosity than normal disk galaxies both in the
field and in clusters as expected from the definition of LCBGs. 

The relationship fit to the low-redshift cluster sample by
\cite{bamford07} was $\log_{10}(r_{50}) = -0.184\times M_B - 3.081$.
Likewise, we fit this same relationship to our data and, as they did,
we fixed the slope of the relationship at a value of $-0.184$.  We find that fitting a line 
of this slope to all cluster galaxies in our sample yields a best-fit intercept of $3.18\pm0.03$, 
which agrees (within the uncertainty) with what Bamford et al. found ($3.20\pm0.02$) 
when they fit a similar line to their sample of intermediate-redshift cluster galaxies.  

We note that if we exclude the LCBGs from our sample and re-fit the line, the derived intercept for the remaining Blue Cloud galaxies in our moderate-redshift \emph{cluster} sample is comparable to  the low-redshift \emph{field} galaxy magnitude-size relationship from \cite{bamford07} with an intercept value of $3.11\pm0.02$. The shift to smaller sizes and more concentrated star formation found by \cite{bamford07} in intermediate clusters is possibly due not to changes in the star formation in  individual disk galaxies, but rather to the increased presence of LCBGs in clusters \citep{c11} that mimics the corresponding increase in number 
seen in the field \citep{guzman97}.

The good agreement between the luminosity-size relationship for cluster and field LCBGs indicates the similarity in these two types of galaxies.  As selected,  LCBGs typically have smaller sizes at a given absolute magnitude than non-LCBG populations.  Although there is a correlation between the size and
luminosity, there is a considerable range in luminosity at any given size.
Due to the extraordinary star formation occurring in some of the galaxies, some of
those small-sized galaxies have magnitudes lying well above the
relationship, thus producing asymmetric scatter about the line.

\subsection{Comparison of Field and Cluster LCBGs}

\begin{figure}[p]
  \epsscale{0.9}
  \plotone{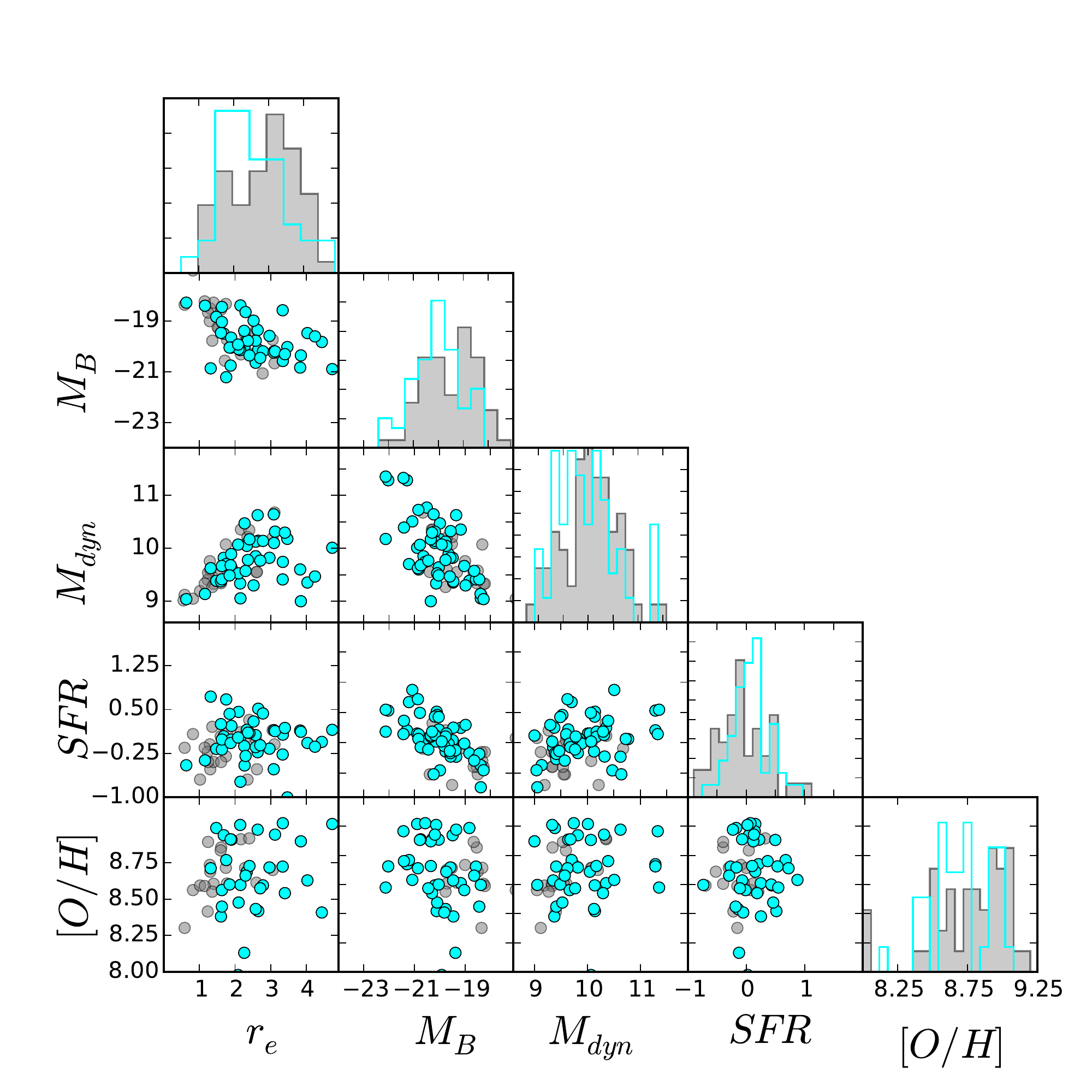}
  \caption{
    \label{fig:allplot}
    Comparison between the size, absolute magnitude, mass, and star
    formation rate for cluster and field LCBGs.  LCBGs that are
    confirmed cluster members from our sample are plotted as turquoise
    circles.  LCBGs from the field (both our sample and from
    \cite{guzman97}) are plotted as grey circles.  No significant
    difference between the field and cluster LCBG populations is
    evident.}
\end{figure}

In Figure~\ref{fig:allplot}, we compare the size, absolute magnitude,
mass, star-formation rates, and metallicity for cluster and field
LCBGs.  For the field sample, we used both the 
field sample from our own observations and the sample from \cite{guzman97}, which
has been selected in an identical manner as our cluster sample. 
We performed a Komolgorov-Smirnov test comparing the field and
cluster samples; in all cases, the null hypothesis that the
distributions were drawn from the same sample could not be rejected.   
We conclude that no difference between the key properties of LCBGs in
our field and cluster samples is apparent.

\subsection{Obscured Star formation in  LCBGs}

\begin{figure}[p]
  \epsscale{0.9}
  \plotone{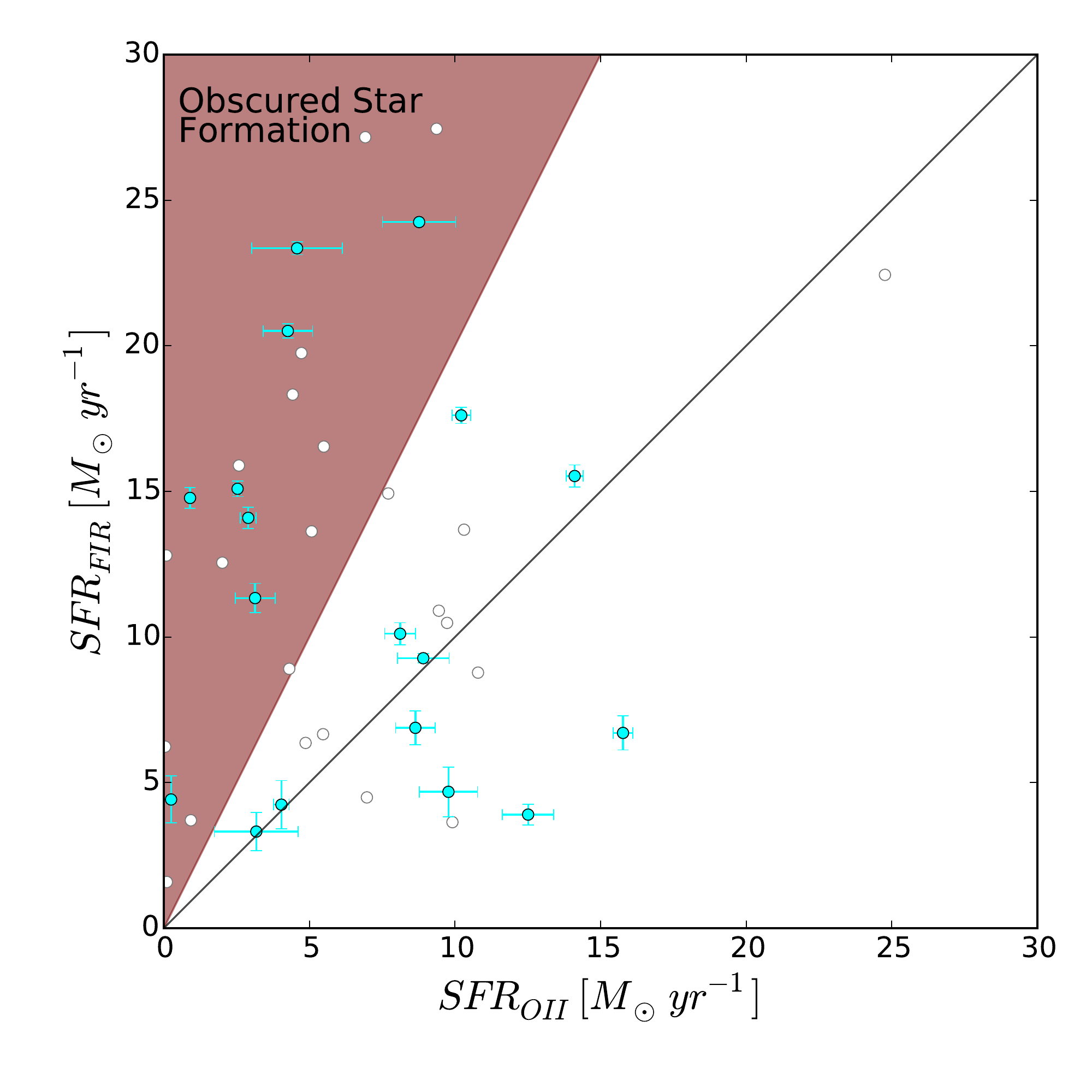}
  \caption{
    \label{fig:sfrir}
    Comparison between star formation measured from
    [\ion{O}{2}]~$\lambda3727$ and $24~\micron$ flux.  Cluster
    (filled-teal circles) and field sources (empty-grey circles) are
    presented.  Approximately $48\%$ of both populations show evidence
    for obscured star formation.   As indicated by the shaded region at top left, we define objects as showing evidence for obscured star formation if their 24 micron-derived SFR exceeds their optically-measured SFR by more than a factor of 2.}
\end{figure}

To explore the fraction of LCBGs with obscured star formation, we
examined the SFR in our LCBGs from our sample with measured
$24~\micron$ fluxes in Figure~\ref{fig:sfrir}.  For the comparison, we
have corrected our optical star-formation rate by a factor of 3 to
account for differences in the assumptions about the IMF (see \S 3.7) .  Although a
number of sources exhibit equivalent SFR between the two
methodologies, an equal number of sources show much higher star formation based on the FIR indicator. We attribute this discrepancy to the presence of obscuring dust in some LCBGs that preferentially absorbs optical radiation and causes the estimated SFR from optical and IR measurements to differ. We define ``obscured''
star-forming galaxies as those having $\mathrm{SFR}_{\mathrm{FIR}} >
2 \times \mathrm{SFR}_{3727}$.  Of the cluster sources measured, 44\% show 
evidence for obscured star formation.

Applying an individual extinction correction to the [\ion{O}{2}]
$\lambda3727$-measured SFR may produce a stronger correlation between
the two star-formation metrics.  For example,
\cite{2012MNRAS.426..330D} found good agreement between H$\alpha$ SFR
and $\mathrm{SFR}_{\mathrm{FIR}}$ in galaxies with detections in both passbands 
only after correcting for the extinction at H$\alpha$ in individual galaxies.  
Unfortunately, we lack the wavelength coverage or data quality to
measure individual extinctions in each of our galaxies.   However, we can make 
some statistical estimates.The galaxies
with the bluest $(U-B)_{o}$ rest-frame colors, strongest $H\beta$ equivalent widths, and strongest
[\ion{O}{3}] $\lambda5006$ equivalent widths show no evidence of obscured star formation, 
whereas galaxies with weaker measurements show a range in the ratio of  
$\mathrm{SFR}_{\mathrm{FIR}}$ to $\mathrm{SFR}_{3727}$.  

We detect no significant difference in the fraction of galaxies exhibiting obscured star formation with
$\sim48\%$ of galaxies in both cluster and field environments showing obscured star
formation.  In addition, we measure no tendency for obscured star formation to be more or less common in objects at greater projected distances from the cluster center,, although the \textit{Spitzer} images do not
cover the full spatial extent of our clusters.   

\subsection{Trends with Cluster Radius}

\begin{figure}[p]
  \epsscale{0.9}
  \plotone{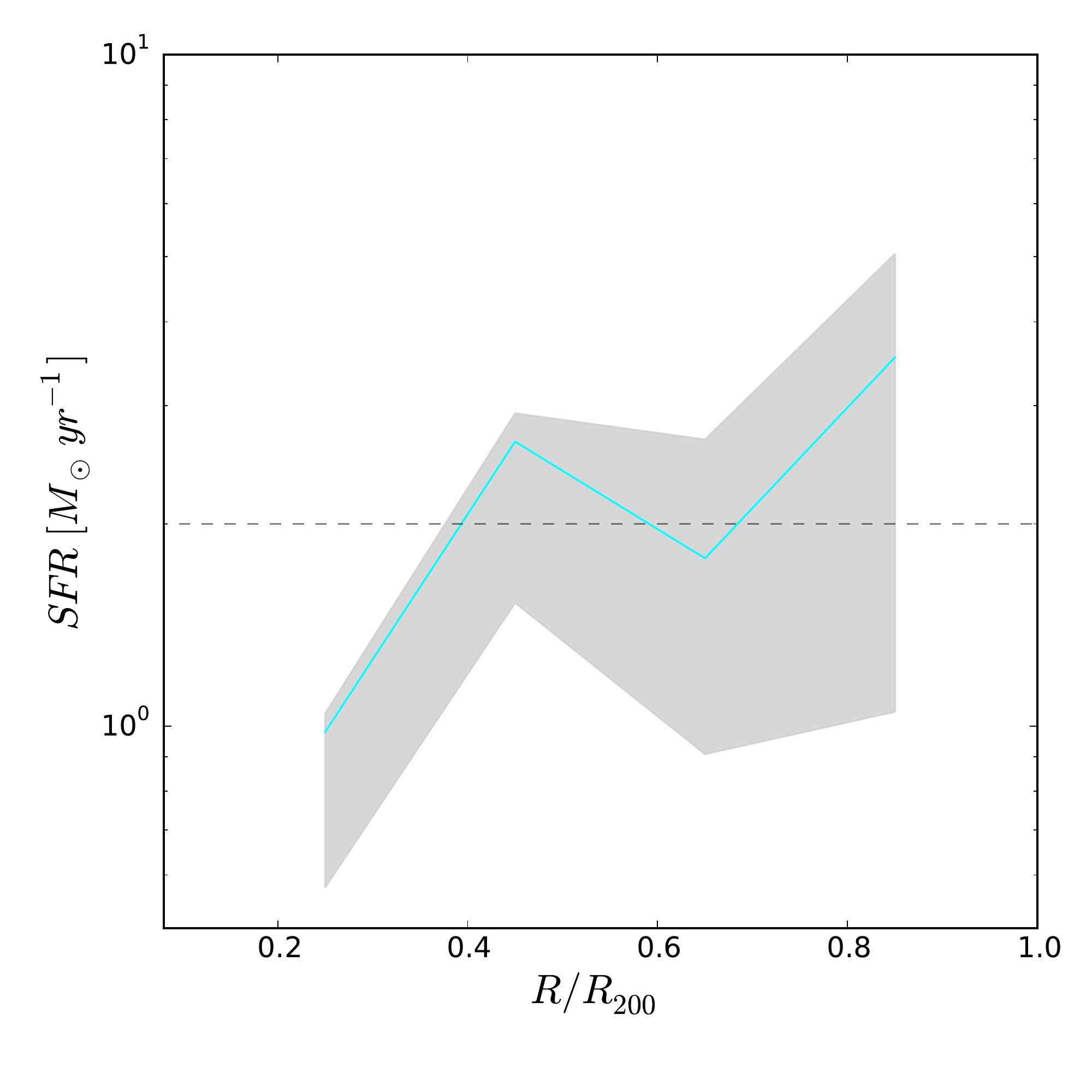}
  \caption{
    \label{fig:vplot}
    Mean star formation rate in LCBGs as a function of cluster radius.  The cluster radius has been normalized by $r_{200}$ for each cluster.  The solid line indicates the mean while  the gray area indicates the upper and lower quartiles.   For comparison, we plot the average star formation rate for field LCBGs from our sample at $z=0.7$ (black, dashed line) .  The LCBGs show evidence for a rapid increase in star formation followed by a decline towards the cluster center.    }
\end{figure}

As we have previously shown \citep{c06, c14}, LCBGs preferentially
avoid the central regions of clusters.  As can be seen in
Figure~\ref{fig:vplot}, strongly-star-forming LCBGs seem to be largely
absent within the cores of the clusters.  The decrease in star-forming
galaxies at small radii appears to begin around $0.5 \ R_{200}$, and within
$0.2 \ R_{200}$ there seems to be a complete lack of strong star-forming
LCBGs.    The distribution of SFR in our sample of intermediate-redshift LCBGs mimics the distribution
seen for low-redshift star bursting galaxies by
\cite{2012MNRAS.427.1252M} with a rise in star formation at larger
radii followed by a decrease towards the cluster center.

Although the SFR as a function of radius for all types of galaxies
shows a strong decrease towards the center of the cluster \citep{2003ApJ...584..210G},  
the SFR for galaxies that show evidence of star formation shows no trend with radius \citep{1997A&A...321...84B,
  2005ApJ...630..206F, 2010A&A...524A..24B}.  However, the overall SFR
in cluster galaxies does tend to be less than that in the field at the
same redshift \citep{2002MNRAS.335...10B, 2005ApJ...630..206F}.  For
LCBGs, we see no difference in SFR between cluster and field for the
population as a whole, but we do observe a  steep decline with cluster
radius.   A similar trend is observed with the $SFR_{IR}$ data as well but 
the limited field of view of the IR data makes it difficult to confirm the trend.
If these sources are recent arrivals to the cluster, the star formation in the galaxies is being  
quickly  quenched as they fall into the cluster.  Similar behavior has been observed for low-redshift star forming galaxies \citep{2012MNRAS.427.1252M}.  

Other than properties associated with star formation, no other
property showed any strong indication of trends with radius.

\section{The Fate of LCBGs in Clusters}
\label{sec:fate}

In \cite{c14}, we argued that cluster LCBGs are galaxies likely
falling into the cluster for the first time based on their observed
spatial and velocity distributions.  In this work, we have shown that
cluster LCBGs as a class are indistinguishable from field LCBGs in
terms of dynamical mass, size, luminosity, metallicity, and
star-formation rate.  When combined with their relative absence from the cluster core,  
this further supports the idea that these galaxies are recent arrivals in the cluster environment.  

Although some LCBGs may plausibly masquerade as low-surface-brightness
disks with centrally-concentrated star formation
\citep{2001ApJ...550L..35B}, deep imaging of LCBGs has revealed that
fewer than $10\%$ of field sources exhibit any extended disk-like
structure \citep{2006ApJ...640L.143N, 2006ApJ...649..129B}.
Regardless, any extended or low-surface-brightness structure will most
likely be stripped away due to various cluster processes \citep{bg06}.
As such, the stars and gas that originally resided within LCBGs are
likely to have either: 1) merged into other cluster galaxies, 2)
dispersed to become part of the intracluster light (ICL), and/or 3)
faded to become dwarf ellipticals.  The following sections explore
these three possible scenarios.

\subsection{Merging of LCBGs}

Based on our previous analysis of projected sky position and velocity
relative to near neighbors, approximately $40\%$ of LCBGs in clusters
appear to be closely associated with another galaxy \citep{c14}.
Previous observations of starburst galaxies in intermediate-redshift
clusters also indicate a high percentage of interacting or merging
systems \citep{1994ApJ...430..121C,2006MNRAS.373..167M,
  2011MNRAS.417.1996J}.  Although the large velocity dispersion among
cluster galaxies is expected to limit the number of mergers which can
occur in the cluster core \citep{1998MNRAS.300..146G}, galaxies
falling in as part of groups may undergo mergers on the outskirts of
clusters \citep{2003astro.ph..5512M}.

An open question is what will happen to these close associations of
galaxies.  After one passage through the cluster core, even small
groups are likely to be tidally disrupted \citep{1994ApJ...433L..61G}.
However, in their simulations \cite{2012ApJ...757...48M} find that
most galaxies will end up merging with a larger galaxy over the
lifetime of the cluster while many of the remaining galaxies will be
tidally disrupted to become part of the ICL.  By $z\sim0.5$, most of
the  tidally disrupted material ends up being accreted onto larger
galaxies.  Detailed modeling of interactions between galaxy pairs in
clusters is required to determine the fraction of LCBGs that will
ultimately merge into other cluster galaxies.

\begin{figure}[p]
  \epsscale{0.9}
  \plotone{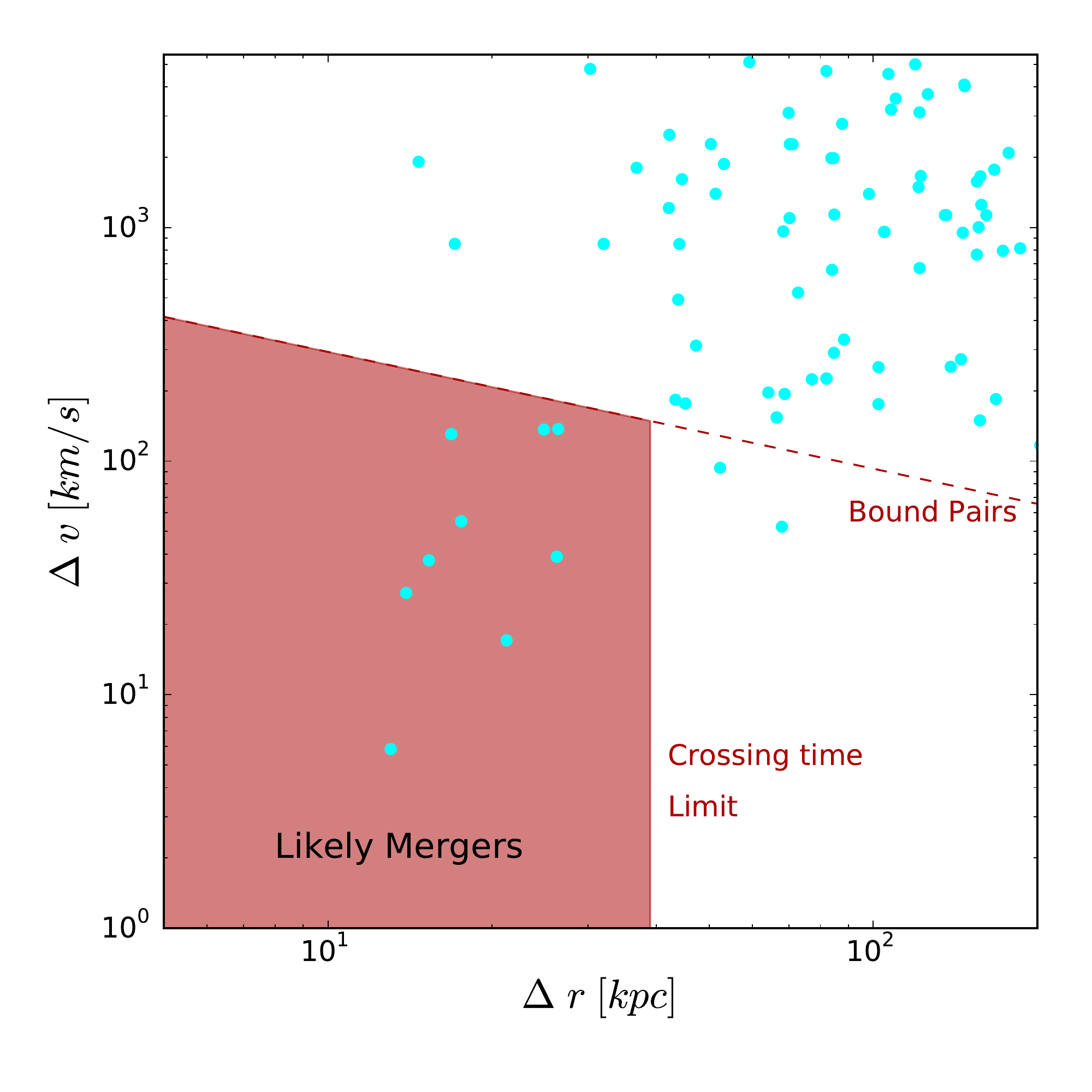}
  \caption{
    \label{fig:neighbor}
    Velocity and projected distance of nearest spectroscopic neighbor
    for each LCBG.  Objects within the pink shaded region are likely
    to form gravitationally-bound pairs that will merge.  The right-hand
    limit to the plot is based on the typical timescale for mergers of
    field galaxies from \cite{2011ApJ...742..103L} and the typical
    crossing time of the cluster, while the upper limit is based on the
    escape velocity from a $M=10^{11} \ M_{\odot}$ companion.  Based on confirmed pairs via
    spectroscopic , we estimate that $10\%$ of LCBGs will undergo a
    merger with their companion. }
\end{figure}

However, we can estimate the number of mergers that are likely to
occur from the velocities and positions of our LCBGs.  For each LCBG,
we plot in Figure~\ref{fig:neighbor} the projected distance and radial
velocity difference between each LCBG and its nearest spectroscopic
neighbor.  From the simulations
of \cite{2011ApJ...742..103L} of the merger timescale for field
galaxies at different separations, we estimate that pairs of galaxies
with separations under 40~kpc are likely to merge within 1~Gyr--the
typical crossing time for these clusters.  After 1~Gyr, the pair would
be expected to be separated by the cluster tidal forces.  This
distance is set as the maximum separation for galaxies likely to
merge.  The other limitation we place on the galaxies is that they
should be gravitationally bound.  The sloping line  in Figure ~\ref{fig:neighbor} defines the
upper-limit of the region in the  plot where
objects  are likely to be gravitationally bound to their
companion, based on their distance and velocity difference if we
assume that the companion has a typical mass of $M=10^{11}
\ M_{\odot}$.   These two considerations
allows us to define an area in Figure~\ref{fig:neighbor} within which objects are likely to be mergers.
Pairs that are near this threshold may still undergo a
merger or experience tidal disruption, depending on their exact
position and motion.  Out of our sample, we estimate that $10\%$ of
sources will undergo a merger.  This number may also be underestimated
due to our spectroscopic incompleteness, as most of the LCBGs in our
sample do not have any spectroscopic confirmation of their nearest projected
neighbor's redshift.  If we include all LCBGs with a bright companion
($R<23$) within 40~kpc, we find that the merger rate could be as high
as 35\%.

\subsection{Disruption of the LCBG Phase}

Several processes serve as potential disruptors of LCBGs in the
cluster environment.  First, ram pressure stripping
\citep{1972ApJ...176....1G} may remove gas from the galaxies.  Second,
the high star-formation rates in LCBGs (relative to their dynamical
masses) can eject a large amount of gas and dust via  winds from star
formation and supernovae.  Third, tidal forces from gravitational
interactions with neighboring galaxies or the overall cluster
potential can also remove gas and stellar
material from the galaxy \citep{1999MNRAS.304..465M}.  

The net result is that some combination of these three disruptive
processes can potentially remove large fractions of the stars and gas
from LCBGs.  For example, simulations of just a  ``galaxy harassment''
scenario by \cite{1999MNRAS.304..465M} found that a typical
high-surface-brightness galaxy\footnote{In \cite{1999MNRAS.304..465M}, high-surface-brightness
galaxies were defined as having a luminosity of $\sim L^*$ and a disc scalelength of $r_d=3 kpc$.   Low-surface-brightness galaxies have a disc scalelength of $r_d=7 kpc$.  }
 would lose 20\% of its stellar mass
over 5~Gyr in a cluster.  For low surface-brightness galaxies, this
percentage increases to 60\%.  The exact percentage of the stripping
will depend on the orbital parameters, with closer approaches to the
cluster center likely to result in greater mass loss
\citep{1998ApJ...495..139M}.  Low-mass galaxies are likely to be
completely disrupted by tidal shocks \citep{2003ApJ...589..752G}.

Furthermore, the high star-formation rate occurring in these systems
is likely to further deplete their gas reservoirs. The timescale for
gas depletion in local LCBGs was found to be on order of 100--200~Myr
by \cite{2005ApJ...624..714G}.  If intermediate-redshift LCBGs have
similar gas reservoirs, they would quickly exhaust their supply of gas
at their observed star-formation rates without any need for stripping
of the cold gas.  Without the accretion of new gas, the transformation
of LCBGs into dE galaxies could easily happen through the process of
strangulation \citep{2008ApJ...672L.103K}.  In addition, the on-going star formation 
is likely to drive strong galactic winds that could potentially deplete as much gas as the
ongoing star formation \citep{2013MNRAS.433..194B}.  If this gas is
swept away by the ICM, it will further hasten the quenching process.
On the other hand, some of the material may remain in the LCBGs due to
confinement from the hot intercluster medium
\citep{1999MNRAS.309..161M}.

Due to the available gas, high star formation rates, and extreme cluster environments, LCBGs will likely
be quenched on very short time scales.   Once star formation ceases, the galaxy is likely to rapidly fade and redden to move out of the LCBG phase \citep{1995ApJ...440L..49K}.  

\subsection{LCBGs as Progenitors of Dwarf Ellipticals}

Regardless of what happens to the stripped stars and gas, LCBGs have
long been proposed as the progenitors of lower-mass galaxies once
their current burst of star formation has faded
\citep{1996ApJ...460L...5G, 1997ApJ...478L..49K}.  The local-group, dE
galaxy NGC~205 is often cited as a possible descendant of LCBGs.
NCG~205 has a velocity dispersion of $42~\mathrm{km~s}^{-1}$, an
effective radius of 2.5~kpc, and an absolute magnitude of $M_B =
-15.0$ \citep{2006AJ....131..332G}.  In addition, the metallicity of
NGC~205 as measured from planetary nebulae is $12+\log(\mathrm{O/H}) = 8.6$
\citep{1995ApJ...445..642R}.  This is very comparable to our sample of
cluster objects which have a median velocity dispersion of
$56~\mathrm{km~s}^{-1}$, a size of 1.8~kpc, and a $12+\log(\mathrm{O/H}) =8.6$.
The median absolute magnitude is $M_B = -19.9$ for cluster LCBGs,
which would represent fading of up to 4.9~mag.  The ultimate
luminosity of a post-starburst system will depend on any underlying
stellar population, the strength of the current burst, and whether the galaxy is quenched.  
For a single stellar population, the expected fading is 5--7~mag after 5-9 Gyrs based
on the models from \cite{bc03}.   For comparison, a model with an exponential-declining star 
formation with e-folding time of 1 Gyr followed by a 100 Myr burst that occurs 5 Gyrs after the galaxy formed and is equal to 10\% of the 
current stellar mass would be expected to fade between 2.5--3~mags over the same period of time.  
Depending
on how extreme these starburst events are, it would be reasonable to assume that a typical 
LCBG could end up as perhaps a slightly smaller version of
NGC~205 once star formation terminates.
   
One longstanding criticism of the hypothesis that LCBGs evolve into dE
is that the gas phase metallicity of the LCBGs was too high for the
average stellar populations found in NGC~205  
\citep{1999ApJ...511..118K} as measured by oxygen abundances 
in planetary nebulae.   Yet, a range of values are found within the 
planetary nebulae in NGC~205  \citep{2014MNRAS.444.1705G}, and
the more appropriate comparison would be with the most metal rich 
planetary nebula in a galaxy, which would reflect the gas phase metallicity
of the last star formation event rather than the average over the history 
of star formation in the galaxy.  Furthermore, typical dE galaxies in the Virgo cluster 
have $\alpha$-element metallicities exhibiting a large scatter 
around solar \citep{2008MNRAS.385.1374M}, which is close to what we measure 
for the intermediate-redshift LCBGs.   

Furthermore, dE galaxies in low redshift clusters show a range of star-formation
histories \citep{2003AJ....125...66C,2008MNRAS.385.1374M,2009MNRAS.396.2133K,2010MNRAS.405..800P} and accretion histories \citep{2001ApJ...559..791C, 2012MNRAS.419.3167S}.
\cite{2014MNRAS.443.3381P} find that dE in Perseus formed from disk
galaxies stripped through harassment and \cite{2008MNRAS.385.1374M}
find that star formation in the dE population in Virgo likely was
truncated as the galaxies were accreted into the cluster over time.
\cite{2009AN....330.1043L} reports that present-epoch dE galaxies form
a morphologically diverse group with $\sim50\%$ having a strong
nucleus, 20\% having a disk, and 10\% having a blue core.  Along
with the different morphologies, the different types are shown to have
a range of stellar ages and distributions within the clusters.  A
single evolutionary path is unlikely, but an important next step will
be to look at the individual morphologies of LCBGs and to simulate how
they will evolve in the cluster environment.

Even though LCBGs form a relatively homogeneous class in these distant clusters of galaxies,
their differing infall trajectories will result in different processes
affecting their subsequent evolution after entering the cluster.
Not only will the cluster environment strip
material from the LCBGs, it will also transform them.
\cite{2009A&A...494..891A} show that fast tidal interactions are very
efficient at stripping the outer parts of galaxies and that bright,
late-type galaxies could easily be transformed into early-type dwarf
galaxies by removing any outer halo or disk of the galaxies.  As dense
systems are likely to be least disrupted \citep{1998ApJ...495..139M},
LCBGs with concentrated star formation are the most likely to survive
in the cluster environment.
Hence, these initially similar galaxies may evolve into a diversity of
morphological types today.  Detailed modeling of each of these
galaxies may help to connect them to their eventual dE counterparts.

\subsection{Number density evolution of LCBGs}

To calculate the number of descendants of LCBGs in present-day galaxy clusters,  we  make the following assumptions about how they will evolve and what their properties are:
\begin{itemize}
\item  10\% of the LCBG sample will merge with other cluster galaxies as measured from our sample. 
\item The stellar-mass-to-dynamical-mass ratio for LCBGs is 0.4 as measured by \cite{2003ApJ...586L..45G}.
\item The star formation of an LCBG is quenched immediately upon entering the cluster.
\item The amount of fading will be given by a \cite{bc03} model that had a star
formation history characterized by declining star formation followed by
a starburst equal to 10\% of its stellar mass that occurs when the galaxy 
is accreted onto the cluster.
\item $20-60\%$ of the stellar mass will be stripped during each passage through the cluster. 
\item The accretion rate of LCBGs at a given redshift will be given by the cluster luminosity function at 
that redshift divided by the visibility period for LCBGs.  
\item The cluster luminosity function of LCBGs evolves linearly between 
those measured above between z=0.5-0.9 to z=0, where it is assumed to have a value of zero.
\item The LCBG phase is visible for between 200-600 Myrs.   
\end{itemize}
The final item is based on the expected lifetime for LCBGs. If
their star formation were immediately quenched, LCBGs would redden
over time so that the typical LCBGs in our sample would no longer meet
the established color criterion within 400~Myr.   At these redshifts and cluster masses,  
the SFR peaks at a radius of 1~Mpc from the
cluster core and almost no star formation is seen within 0.25~Mpc of
the cluster core.  A typical object within the cluster would take
600~Myr to cross this distance.  Accordingly, we take this as the
upper limit for the visibility time for an LCBG.  For a lower limit, 
we use the estimate of 200~Myr from \cite{2005ApJ...624..714G} based 
on the amount of time it would take for the LCBGs to exhaust their gas reservoir. 

\begin{figure}[p]
  \epsscale{0.9}
  \plotone{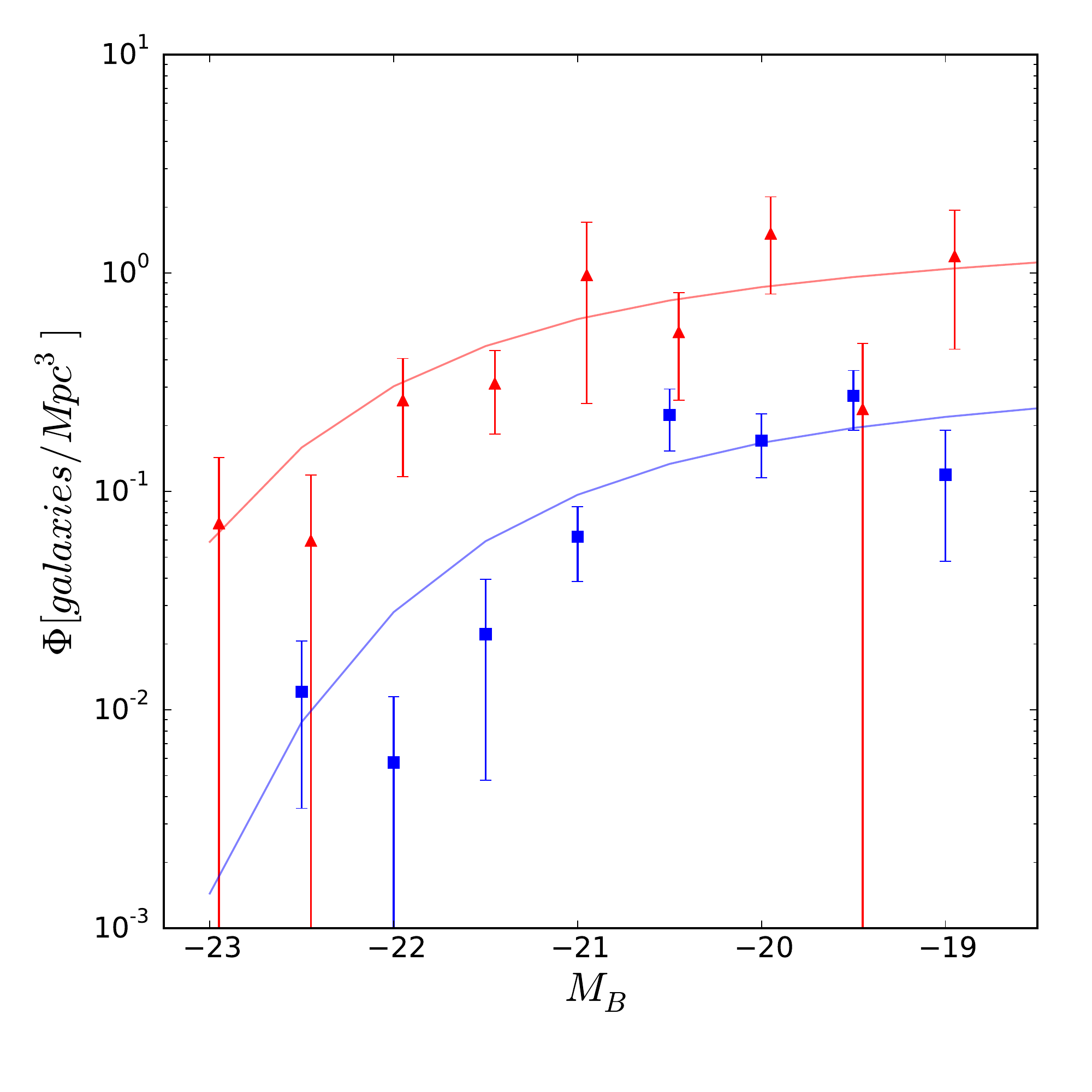}
  \caption{
    \label{fig:lf}
    The $z=0.55$ (blue squares) and $z=0.83$ (red triangles) luminosity
    functions for LCBGs.  To both luminosity functions, we have fit
    Schechter functions with a fixed slope of $\alpha=1.1$.   The cluster LCBG functions
     show the similar decrease in number density as seen in the field.  The z=0.83 luminosity 
     function has been offset by 0.05 mag for clarity.}
\end{figure}

Based on our sample, we can calculate the evolution in the cluster luminosity function.  Figure~\ref{fig:lf} compares the $0.7<z<0.9$ and $z=0.55$ luminosity
functions of cluster LCBGs.  The luminosity functions were calculated
by summing over all LCBGs in the cluster within $R_{200}$ and correcting for
spectroscopic incompleteness in the same manner as in \cite{c11}.    The luminosity
function for each cluster was then normalized by the volume of the cluster as estimated
from the $R_{200}$ values given in Table 1 of \cite{c14}.   Finally, we averaged the luminosity functions
into the two redshift bins.  For reference, the average volume of our z=0.55 (z=0.83) redshift  clusters is 
$74 \ Mpc^3$ ($16 \ Mpc^3$ ).  The values for the observed luminosity functions are 
given in Table~\ref{tab:lf}.  For both luminosity functions, we have fit Schechter functions with a fixed 
slope of $\alpha=1.1$ .  At $z=0.55$, the luminosity function has a value of $M_B = -21.30$ and 
$\phi_* = 0.217$  galaxies/Mpc$^3$; at $z=0.83$, $M_B = -21.97$ and 
$\phi_* = 0.92$ galaxies/Mpc$^3$.  As can be seen, the number density and luminosity of LCBGs rapidly evolve
between $z=0.83$ and $z=0.55$, an effect reported previously
\citep{phillips97, c11}. 

\begin{deluxetable}{rrrrrrrrrrr}
\tabletypesize{\tiny}\tablewidth{0pc}
\tablecaption{LCBG Luminosity Functions}
\tablehead{
\colhead{$M_B$} & 
\colhead{$\phi$ at z=0.55} & 
\colhead{$\phi$ at z=0.83} \\
\colhead{} & 
\colhead{galaxies/Mpc$^3$} & 
\colhead{galaxies/Mpc$^3$} \\
}
\startdata 
\label{tab:lf}
-23.00 & $0.000 \pm 0.000$ & $0.071 \pm 0.071$ \\
-22.50 & $0.012 \pm 0.009$ & $0.060 \pm 0.060$ \\
-22.00 & $0.006 \pm 0.006$ & $0.486 \pm 0.243$ \\
-21.50 & $0.022 \pm 0.017$ & $0.427 \pm 0.180$ \\
-21.00 & $0.062 \pm 0.023$ & $1.013 \pm 0.463$ \\
-20.50 & $0.224 \pm 0.071$ & $0.611 \pm 0.327$ \\
-20.00 & $0.171 \pm 0.055$ & $1.256 \pm 0.452$ \\
-19.50 & $0.274 \pm 0.084$ & $0.092 \pm 0.092$ \\
-19.00 & $0.119 \pm 0.071$ & $1.230 \pm 0.541$ \\
\enddata 
\end{deluxetable}

\begin{figure}[p]
  \epsscale{0.9}
  \plotone{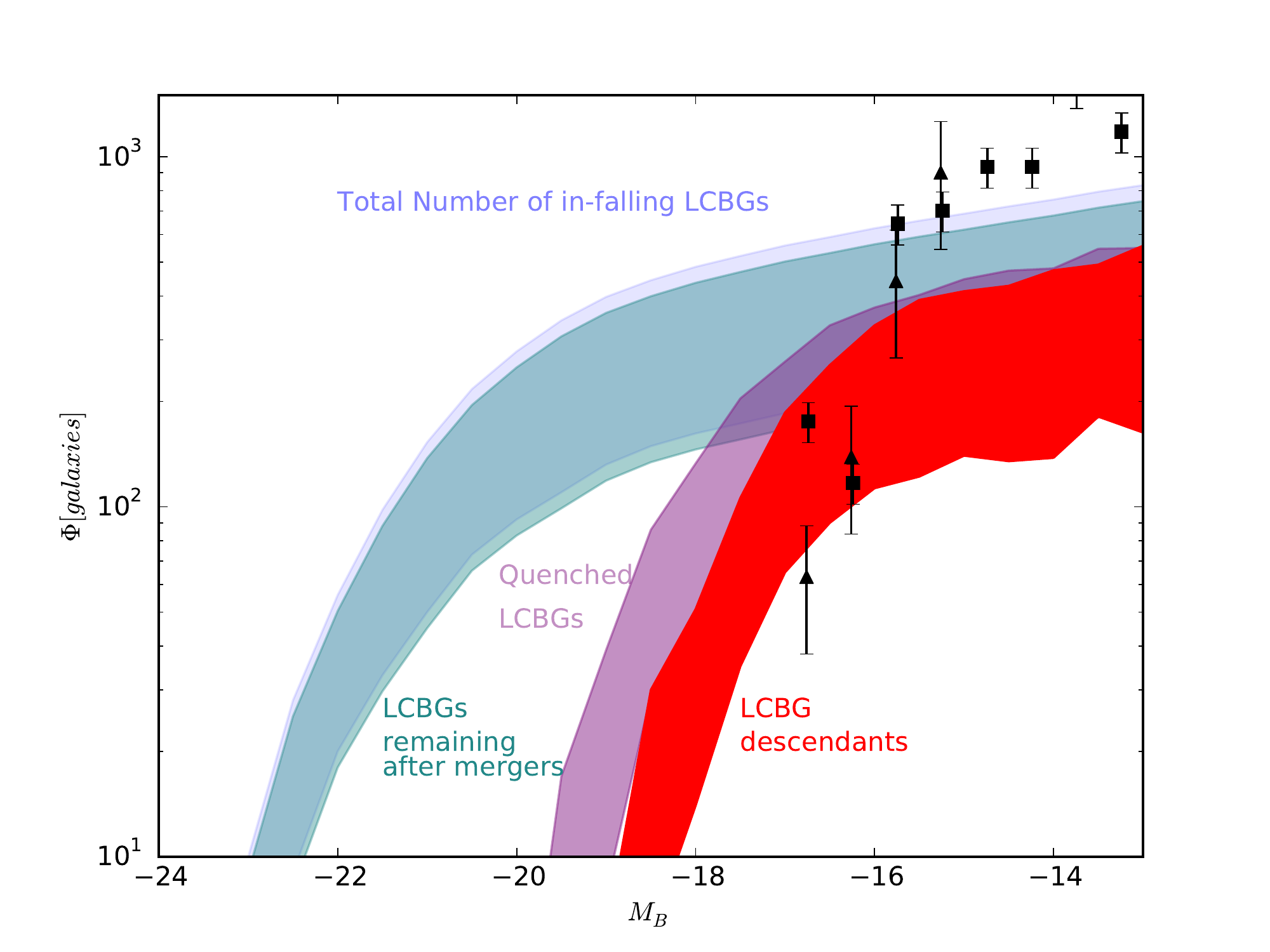}
  \caption{
    \label{fig:sim}
    Predicted evolution of the number density of LCBGs in a massive
    galaxy cluster.  The gray curve represents the total number
    of LCBGs accreted onto the cluster since $z=1.0$.  The width of
    the curve represents two assumptions about the amount of time the
    LCBG phase is visible, with the upper limit representing a
    visibility window of 200~Myr and the lower limit corresponding to
    600~Myr. The teal curve represents the number of LCBGs
    remaining after 10\% of LCBGs merge with other galaxies.  The purple
    curve represents the distribution of LCBGs after their star
    formation is quenched and they begin to fade.  The descendants of
    LCBGs -- after accounting for mergers, fading, and stripping from the
    cluster environment -- lie within the red region.  For comparison, the
    luminosity function for dwarf elliptical galaxies in Fornax from
    \cite{1988AJ.....96.1520F} and Coma from  \cite{1993AJ....106.2197T} normalized to the mass and area of 
    our clusters at $z=0$ is plotted as black squares and triangles, respectively.  It is likely that
    approximately $50\pm20\%$ of dE galaxies seen in present day
    clusters underwent an LCBG phase since $z=1$. }
\end{figure}

Using  the observed evolution in the LCBG luminosity function, we
simulate a cluster accreting LCBGs between $z=1.0$ and $z=0.0$.  Once
an LCBG is accreted,  it evolves based on the assumptions
outlined above.  The resulting luminosity functions for these sources
are presented as the red region in Figure~\ref{fig:sim}.
Overall, the accretion of LCBGs into galaxy clusters would result
between 600--1800 objects with $M_B<-10$ being added to the cluster.
These numbers are similar to the number of dE found in low-redshift
clusters \citep{1988AJ.....96.1520F, 1997PASP..109.1377S}.  For
comparison, we plot the luminosity function for dEs in the Fornax
Cluster from \cite{1988AJ.....96.1520F} and Coma Cluster from \cite{1993AJ....106.2197T}.  We have normalized both
luminosity functions to the expected $z=0$ mass ($4.9\times10^{15} \ M_{\odot}$) of our
intermediate-redshift cluster  based on the predictions from
\cite{2002ApJ...568...52W} and corrected the area of the original 
surveys to a value of $R_{200}$.

For the assumptions outlined above, we estimate that LCBGs accreted
between z=1.0 and 0 account for 38\% of dE galaxies brighter than  $M_B < -14.5$.    Changes to the assumptions about the stripping in  galaxies will result in a maximum change of 10\% in the fraction of dE descendants.   An increase (decrease) in the amount of fading by 1 mag will change the percentage of dEs to 23\% (50\%).  Finally, the largest effect on the    percentage of bright dEs is the visibility of the LCBG phase.   For  the shorter period of 200 Myr, LCBGs could account for up to 77\% of dEs;  for 600 Myrs, LCBGs would only account of $25\%$ of dEs.  
  
The fraction of dE galaxies in present-day clusters which were once LCBGs remains uncertain. The simple assumptions  made about their evolution do not match in detail the shape of the observed low redshift luminosity functions of dwarf ellipticals as can be seen in Figure \ref{fig:sim}.   However, even at the lowest estimates, a significant fraction of dEs in local clusters would have gone through an LCBG phase at intermediate redshifts.   Better limits on how these galaxies have evolved require more detailed investigation of the evolution of individual galaxies, taking a holistic look at the properties of local dE galaxies, and improving measurements of their luminosity functions.Ó

\section{Conclusion}

In this work, we have reported on the rest-frame properties of LCBGs,
an extreme star-bursting population seen at intermediate redshifts.
Based on ground- and space-based imaging and spectroscopy, we have
measured the size, luminosity, star-formation rate, metallicity, and
dynamical mass of the LCBGs.  Based on these measurements, we have
observed:

\begin{enumerate}
\item Much of the evolution in the size-luminosity relationship for
  blue cluster galaxies appears to be driven by the inclusion of LCBGs
  at intermediate redshift.  No evolution is observed in this relationship if
  LCBGs are excluded from the fit. 
\item We detect no significant differences in the size, mass, luminosity, star formation rate
 or metallicity  of
  field and cluster LCBGs.  To the degree we can discern, all of the
  key characteristics appear similar between the two populations.
\item $ 35\%$ of cluster LCBGs show evidence for obscured star
formation with no strong trends found with either LCBG properties or
cluster position.
\item The star-formation rate of cluster LCBGs decreases towards
  the core of the cluster.  Almost no LCBGs are observed in the core
  of these clusters and the star formation appears to peak beyond  $0.5 R_{200}$.
\item Of the LCBGs currently falling into the cluster, 10\% are
  likely to merge with a nearby neighbor as they pass through the
  cluster.
\item The size, mass, and metallicity of LCBGs at intermediate
  redshift are very similar to the properties of present-day dwarf
  elliptical galaxies.  Galaxy evolution models predict that if their
  star formation were quenched, the LCBG population would evolve in
  magnitude and color so as to produce galaxies that closely resemble
  today's dwarf elliptical population.
\item Based on assumptions about their evolution and accretion rate,
  LCBGs seen at intermediate redshifts ($z<1$) could account for
  30--75\% of all dwarf ellipticals seen in clusters today.
\end{enumerate}

An important next step will be to trace the evolution of these sources
through realistic simulations of the cluster environment accounting
for the physics of both the stellar and gas components of these
galaxies. As the distribution and properties of the LCBG population
matches the population of dwarf elliptical galaxies in clusters, it
will be important to determine which physical interactions with the
cluster environment are the predominant mechanisms in transforming
LCBGs into dwarf elliptical galaxies.  Exploring how the LCBG
population fits into the evolution of other cluster populations and how the 
overall cluster population changes with time will be critical to understanding
the transformation of galaxies in clusters.

\acknowledgements

We thank the referee for the careful reading of our manuscript
and the constructive criticism that improved our paper.
S.M.C. acknowledges the South African Astronomical Observatory and
the National Research Foundation of South Africa for support
during this project. M.A.B. acknowledges support from NSF
grant AST-1009471.

This research made use of Astropy, a community-developed core Python package for Astronomy \citep{2013A&A...558A..33A}.   This research has made use of the NASA/IPAC Extragalactic Database (NED) and NASA/IPAC
Infrared Science Archive which are operated by the Jet Propulsion Laboratory, California Institute of Technology, under contract with the National Aeronautics and Space Administration.  

Based on observations made with the NASA/ESA Hubble Space Telescope, obtained from the data archive at the Space Telescope Science Institute. STScI is operated by the Association of Universities for Research in Astronomy, Inc. under NASA contract NAS 5-26555.

The WIYN Observatory is a joint facility of the University of Wisconsin-Madison, Indiana University, the National Optical Astronomy Observatory and the University of Missouri.

Our findings are based on observations acquired at the W. M. Keck Observatory, which operates as a scientific partnership among the California Institute of Technology, the University of California and the National Aeronautics and Space Administration.   The generous financial support of the W. M. Keck Foundation made the Observatory possible.

The authors wish to recognize and acknowledge the very significant cultural role and reverence that the summit of Maunakea has always had within the indigenous Hawaiian community. We consider ourselves fortunate to have had the opportunity to conduct observations from this mountain.

{\it Facilities:} \facility{WMKO, WIYN}

\bibliography{ref}

\end{document}